\documentclass[iop]{emulateapj}
\usepackage[left]{lineno}
\usepackage{graphicx, subfigure}
\usepackage{natbib}
\usepackage{xspace}

\newcommand{\Msun}{$M_{\odot}$\xspace}
\newcommand{\Rsun}{$R_{\odot}$\xspace}
\newcommand{\dmu}{pc~cm$^{-3}$}
\newcommand{\us}{$\mu$s\xspace}

\newcommand{\fermi}{\textit{Fermi}\xspace}
\newcommand{\xmm}{\textit{XMM-Newton}\xspace}

\newcommand{\efluxu}{$\times 10^{-12}$ erg cm$^{-2}$ s$^{-1}$}
\newcommand{\fluxu}{$\times 10^{-9}$  ph cm$^{-2}$ s$^{-1}$}

\newcommand{\msp}{J1400$-$1431\xspace}
\newcommand{\psr}{PSR~J1400$-$1431\xspace}
\newcommand{\Rc}{\ensuremath{R_{\mathrm c}}}

\newcommand{\Teff}{\ensuremath{T_{\rm eff}}}

\begin{document}

\title{A multi-wavelength study of nearby millisecond pulsar \psr: \\ improved astrometry \& an optical detection of its cool white dwarf companion}
\author{J.~K.~Swiggum\altaffilmark{1},
	D.~L.~Kaplan\altaffilmark{1},
	M.~A.~McLaughlin\altaffilmark{2,3},
	D.~R.~Lorimer\altaffilmark{2,3},
	S.~Bogdanov\altaffilmark{4},
	P.~S.~Ray\altaffilmark{5},
	R.~Lynch\altaffilmark{6,3},
	P.~Gentile\altaffilmark{2,3},
	R.~Rosen\altaffilmark{7},
	S.~A.~Heatherly\altaffilmark{6},
	B.~N.~Barlow\altaffilmark{8},
	R.~J. Hegedus\altaffilmark{8},
	A.~Vasquez Soto\altaffilmark{8},
	P.~Clancy\altaffilmark{8},
	V.~I. Kondratiev\altaffilmark{9,10},
	K.~Stovall\altaffilmark{11,12},
	A.~Istrate\altaffilmark{1},
	B.~Penprase\altaffilmark{13},
	E.~C.~Bellm\altaffilmark{14}
	} 

\altaffiltext{1}{Center for Gravitation, Cosmology and Astrophysics, Department of Physics, University of Wisconsin--Milwaukee, P.O. Box 413, Milwaukee, WI 53201, USA}
\altaffiltext{2}{Department of Physics and Center for Gravitational Waves and Cosmology, West Virginia University, White Hall, Morgantown, WV 26506, USA}
\altaffiltext{3}{Center for Gravitational Waves and Cosmology, West Virginia University, Chestnut Ridge Research Building, Morgantown, WV 26505}
\altaffiltext{4}{Columbia Astrophysics Laboratory, Columbia University, New York, NY 10027, USA}
\altaffiltext{5}{Space Science Division, Naval Research Laboratory, Washington, DC 20375-5352, USA}
\altaffiltext{6}{Green Bank Observatory, PO Box 2, Green Bank, WV, 24944, USA}
\altaffiltext{7}{NRAO, 520 Edgemont Road, Charlottesville, VA 22903, USA}
\altaffiltext{8}{One University Parkway, Department of Physics, High Point University, High Point, NC  27268}
\altaffiltext{9}{ASTRON, The Netherlands Institute for Radio Astronomy, Postbus 2, 7990 AA, Dwingeloo, The Netherlands}
\altaffiltext{10}{Astro Space Centre, Lebedev Physical Institute, RussianAcademy of Sciences, Profsoyuznaya Str. 84/32,Moscow 117997, Russia}
\altaffiltext{11}{NRAO, PO Box 0, Socorro, NM 87801, USA}
\altaffiltext{12}{Dept. of Physics and Astronomy, Univ. of New Mexico, NM 87131, USA}
\altaffiltext{13}{Department of Physics \& Astronomy, Pomona College, 610 N. College Ave., Claremont, CA 91711, USA}
\altaffiltext{14}{Department of Astronomy, University of Washington, Seattle, WA 98195}

\begin{abstract}
In 2012, five high school students involved in the Pulsar Search Collaboratory discovered the millisecond pulsar \psr and initial timing parameters were published in \cite{rsm+13} a year later. Since then, we have obtained a phase-connected timing solution spanning five years, resolving a significant position discrepancy and measuring $\dot{P}$, proper motion, parallax, and a monotonic slope in dispersion measure over time. Due to \psr's proximity and significant proper motion, we use the Shklovskii effect and other priors to determine a 95\% confidence interval for \psr's distance,  $d=270^{+130}_{-80}$~pc. With an improved timing position, we present the first detection of the pulsar's low-mass white dwarf (WD) companion using the Goodman Spectrograph on the 4.1-m SOAR telescope. Deeper imaging suggests that it is a cool DA-type WD with $T_{\rm eff}=3000\pm100$~K and $R/R_\odot=(2.19\pm0.03)\times10^{-2}\,(d/270~{\rm pc})$. We show a convincing association between \psr and a $\gamma$-ray point source, 3FGL J1400.5$-$1437, but only weak (3.3-$\sigma$) evidence of pulsations after folding $\gamma$-ray photons using our radio timing model. We detect an X-ray counterpart with \xmm but the measured X-ray luminosity ($1\times10^{29}$~ergs~s$^{-1}$) makes \psr the least X-ray luminous rotation-powered millisecond pulsar (MSP) detected to date. Together, our findings present a consistent picture of a nearby ($d\approx230$~pc) MSP in a 9.5~day orbit around a cool, $\sim$0.3~\Msun WD companion, with orbital inclination, $i\gtrsim60^\circ$.
\end{abstract}


\section{Introduction}

\psr is a 3.08~ms radio pulsar discovered by Pulsar Search Collaboratory (PSC) students \citep{rsm+13} in a portion of the Green Bank 350~MHz Drift Scan Survey \citep{blr+13, lbr+13}. With a dispersion measure (DM) of 4.9~\dmu, it is one of only five millisecond pulsars (MSPs) with ${\rm DM}<5$~\dmu. Since DM provides a measure of the electron content along the line of sight, it can be used as a proxy for distance, given Galactic electron density models \citep[e.g.][]{tc+93,cl+02,ymw+17}. \cite{ymw+17} describe the most recent electron density model, which predicts that \msp has a distance of only 350~pc.

Nearby MSPs allow high-precision measurements of astrometric parameters like proper motion and, in some cases, parallax through pulsar timing. The latter involves detecting the curvature of incoming wavefronts \--- a signature only found in timing residuals for a handful of nearby MSPs close to the ecliptic plane \citep{ktr+94, cfw+94, sbm+97, wdk+00, jbv+03, hbo+04, lkd+04, sns+05,rhc+16,dcl+16,mnf+16}. However, parallax has also been detected using very long baseline interferometry (VLBI) follow-up in many other cases \citep{bbg+02,cbv+09}. Together, distance and DM provide an average measure of free electrons along the line of sight to the pulsar \citep{tbm+99, lkn+06}; combined with proper motion, transverse velocities can be derived to study an underlying distribution for MSPs \citep{tsb+99} and compare it to velocity distributions for other sub-populations. Underlying velocity distributions provide estimates for pulsars' natal kicks from the supernova explosions that created them \citep{hll+05}.

Because of its proximity and brightness, \msp was considered for inclusion in pulsar timing arrays \citep[PTAs; e.g.][]{dfg+13,abb+15}, but was dropped due to inconsistent detectability at 820~MHz and higher observing frequencies. \cite{rsm+13} hypothesized that unreliable detections at higher frequencies were likely due to \msp's particularly steep spectrum. 

\begin{figure*}[ht!]
\begin{center}
\includegraphics[width=0.9\textwidth]{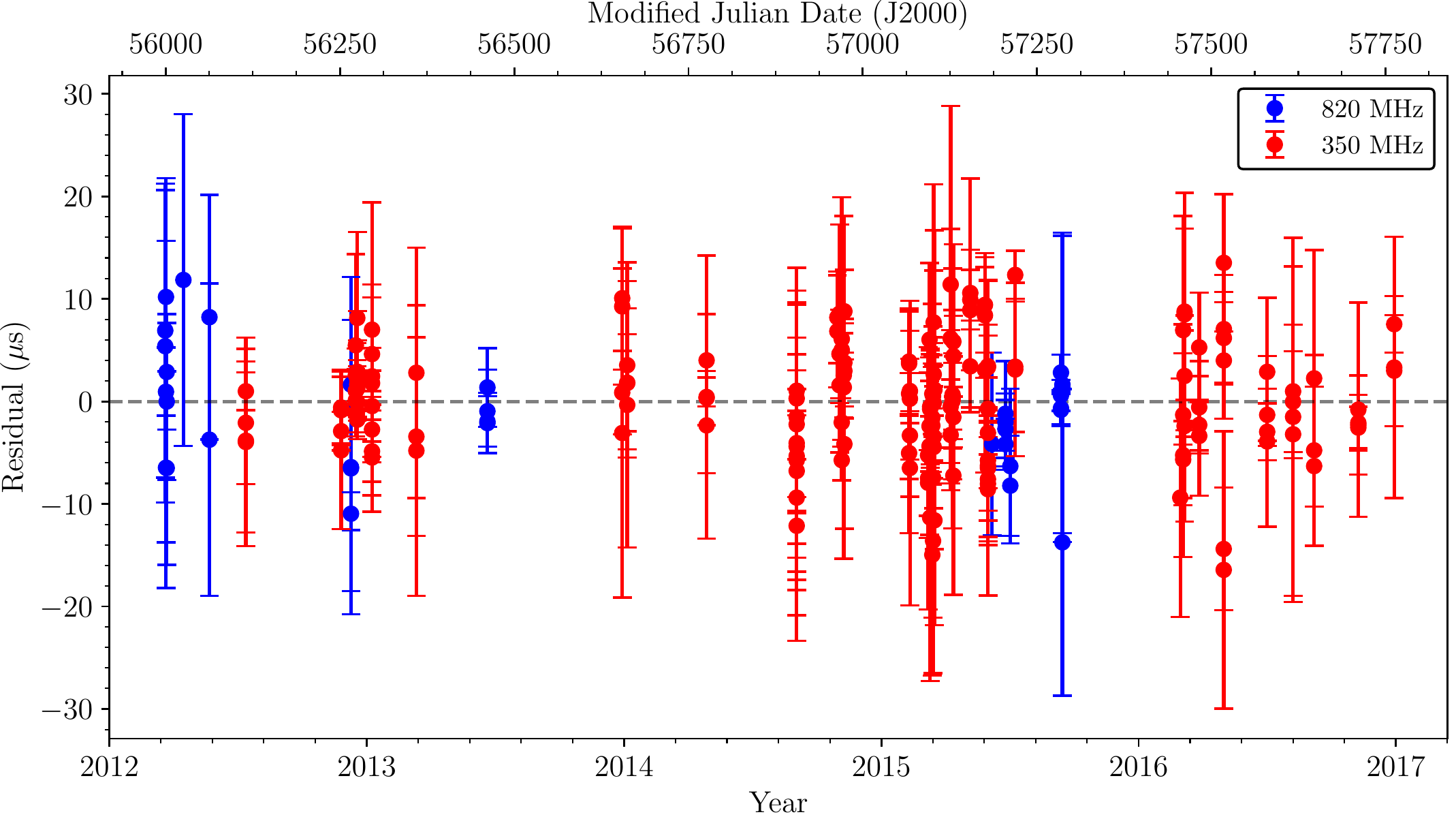}
\caption{Timing residuals in microseconds for \msp, showing observations at 350~MHz (red) and 820~MHz (blue), respectively.}
\label{fig:res}
\end{center}
\end{figure*}

We used a novel drift-scan technique to improve localization for this pulsar (see further discussion in Gentile \& Swiggum, in prep.), finding a position that differed by $6.7\arcmin$ from that published in \cite{rsm+13}. This difference is larger than the formal uncertainty, but since the previous timing solution was based on less than one year of timing data, it is subject to significant covariance between position and spin-down parameters. The offset also undoubtedly played a significant role in early detectability issues at higher frequencies. In this paper, we present an improved, phase-connected timing solution for \msp with pulse times of arrival (TOAs) spanning five years, including those published in \cite{rsm+13}. The significantly longer timing baseline compared to that of the previous study rules out any covariance between fits for position and spin-down. 

In \S\ref{sec:time}, we provide a detailed description of our full radio timing analysis, including measurements of proper motion, a linear slope in DM over time, and first and second Laplace parameters (effectively the orbital eccentricity). We have also developed a posterior probability distribution for \msp's distance based on a timing parallax fit, combined with several other priors. 

Nearby MSPs are also good candidates for multi-wavelength follow-up. In \S\ref{sec:optical}, we describe our observing campaign and photometry analysis using the Keck Low-Resolution Imaging Spectrometer (LRIS) and the Southern Astrophysical Research (SOAR) optical telescopes to image \msp's white dwarf (WD) companion. \psr has spin and orbital parameters similar to other low-mass binary pulsars (LMBPs) \--- namely, its short spin period ($P<10$~ms), small eccentricity ($e<10^{-3}$), and a minimum companion mass, $m_{\rm c, min}=0.26$~\Msun, which falls in a typical range for LMBPs, $0.15$~\Msun~$<m_{\rm c}<0.4$~\Msun. These systems are thought to evolve from a neutron star accreting material from a low-mass star in its giant phase. Stable mass transfer causes the neutron star to spin faster, while its companion (provided $m_{\rm c}\lesssim1.6$~\Msun) does not undergo helium ignition in its core, resulting in a binary system containing a MSP and a low-mass He-core WD \citep{pk+94}.

In \S\ref{sec:gamma} and \S\ref{sec:xray} respectively, we describe $\gamma$-ray and X-ray detections, which also help constrain the pulsar's distance and spin period derivative ($\dot{P}$), taking into account respective emission efficiencies \citep[e.g.][]{gsl+16,becker+09,pb+15}. We synthesize and discuss the collected information from multi-wavelength follow-up in \S\ref{sec:discuss} and summarize our conclusions in \S\ref{sec:conclude}.

\section{Radio Observations \& Timing Analysis}\label{sec:time}

\begin{deluxetable*}{ccccccc}
 \tablewidth{0pt}
\tablecaption{Details of Observing Modes Used for \psr\label{tab:detail}}
\tablehead{
  \colhead{Center Frequency} & \colhead{Bandwidth} &
  \colhead{N$_{\rm channels}$} & \colhead{t$_{\rm sample}$}
  & \colhead{Observing Mode} & \colhead{GUPPI Offset\tablenotemark{a}} & \colhead{N$_{\rm TOA}$} \\
\colhead{(MHz)} & \colhead{(MHz)} & & \colhead{(\us)} & & \colhead{(\us)} & }
\startdata
350 & 100 & 2048 & 81.92 & Incoherent & 40.96 & \phn52 \\
350 & 100 & 4096 & 81.92 & Incoherent & 81.92 & 101 \\
350 & 100 & 128 & \phn1.28 & Coherent Fold & \phn7.68 & \phn17 \\
820 & 200 & 2048 & 81.92 & Incoherent & 20.48 & \phn16 \\
820 & 200 & 128 & \phn0.64 & Coherent Fold & \phn3.84 & \phn17 \\
\enddata
\tablenotetext{a}{Mode-dependent instrumental timing offsets used for \psr.}
\end{deluxetable*}

In order to improve upon the preliminary timing solution published in \cite{rsm+13}, we include those data here, but have reprocessed them according to the procedure described below. All timing observations were conducted with the Robert C. Byrd Green Bank Telescope (GBT) at either 350 or 820~MHz using the Green Bank Ultimate Pulsar Processing Instrument \citep[GUPPI;][]{drd+08} with 100 or 200~MHz bandwidth, respectively, and sampled every 81.92~\us. Since \msp has not been observed as part of a dedicated timing proposal since 2013, many of the more recent TOAs come from using the pulsar to conduct test scans before Green Bank North Celestial Cap \citep[GBNCC; ][]{slr+14} survey observations. Because of this, the set-up/observing parameters changed slightly for different groups of TOAs, so we noted these changes and carefully accounted for any resulting systematics (e.g. GUPPI offsets; see Table \ref{tab:detail}). Also, because of its frequent use as a test source, many scans were taken using incoherent search-mode (rather than coherent fold-mode), resulting in relatively coarse time sampling for MSP monitoring.

We identified the highest signal-to-noise ratio detections at each observing frequency after folding data with the correct spin period and DM at each epoch, then fit three Gaussians to the corresponding pulse profiles to generate noiseless standard profiles. Standard profiles were aligned using {\tt pas} from {\sc PSRCHIVE}\footnote{http://psrchive.sourceforge.net/} \citep{hsm+04}.

\begin{deluxetable}{lc}
\tablewidth{0pt}
\tablecaption{Measured and derived timing parameters for \psr\label{tab:1400par}}
\tablehead{
  \colhead{Parameter} & \colhead{Value}
  }
\startdata
  \cutinhead{Spin \& Astrometric Parameters}
Ecliptic Longitude (J2000)\dotfill & 213.11368082(8) \\
Ecliptic Latitude (J2000)\dotfill & $-$2.1064331(18) \\
Proper Motion in Ecliptic Lon. (mas/yr)\dotfill & 34.75(19) \\
Proper Motion in Ecliptic Lat. (mas/yr)\dotfill & $-$46(6) \\
Parallax (mas)\dotfill & 3.6(11) \\
Spin Period (s)\dotfill & 0.00308423326039194(8) \\
Period Derivative (s/s)\dotfill & 7.2333(15)$\times 10^{-21}$ \\
Intrinsic Period Derivative (s/s)\dotfill & $<2.2\times 10^{-21}$ \\
Dispersion Measure (\dmu)\dotfill & 4.93258(3) \\
$d\,{\rm DM}/dt$ (\dmu~yr$^{-1}$)\dotfill & 1.8(3)$\times 10^{-4}$ \\
Reference Epoch (MJD)\dotfill & 56960.0 \\
Span of Timing Data (MJD)\dotfill & $56006$-$57751$ \\
Number of TOAs\dotfill & 203 \\
RMS Residual (\us)\dotfill & 4.06 \\
EFAC\dotfill & 1.8 \\
\cutinhead{Binary Parameters\tablenotemark{a}}
Orbital Period (days)\dotfill & 9.5474676743(19) \\
Projected Semi-major Axis (lt-s)\dotfill & 8.4212530(6) \\
Epoch of Ascending Node (MJD)\dotfill & 56958.38397673(9) \\
First Laplace Parameter\dotfill & 2.8(12)$\times 10^{-7}$ \\
Second Laplace Parameter\dotfill & 4.8(14)$\times 10^{-7}$ \\
\cutinhead{Derived Parameters}
Right Ascension (J2000)\dotfill & 14:00:37.00370(15) \\
Declination (J2000)\dotfill & $-$14:31:47.0422(6) \\
Orbital Eccentricity\dotfill & 5.5(14)$\times 10^{-7}$ \\
Surface Magnetic Field (10$^{7}$ Gauss)\dotfill & $<8.3$ \\
Spin-down Luminosity (10$^{33}$ erg/s)\dotfill & $<3.0$ \\
Characteristic Age (Gyr)\dotfill & $>22$ \\
Total Proper Motion (mas/yr)\dotfill & 57(5) \\
Transverse Velocity\tablenotemark{b} (km/s)\dotfill & 76(20) \\
Shklovskii Period Derivative\tablenotemark{b} (s/s)\dotfill & 7(2)$\times 10^{-21}$ \\
Mass Function (\Msun)\dotfill & 0.0070345527(14) \\
Minimum Companion Mass\tablenotemark{c} (\Msun)\dotfill & 0.26\\
\enddata
\tablenotetext{a}{Using the ELL1 binary timing model.}
\tablenotetext{b}{Computed using the distance derived from the timing
  parallax measurement with no correction.}
\tablenotetext{c}{Calculated assuming a pulsar mass, m$_p=1.35$~\Msun.}
\tablecomments{Quantities are listed with 68\% (1-$\sigma$) uncertainties on the last digit in
  parentheses. The intrinsic spin-down ($\dot{P}_{\rm int}$) is constrained by $\dot{P}_{\rm Shklov}=5\times10^{-21}$; upper/lower limits
  on other derived parameters come from $\dot{P}_{\rm int}$, assuming the pulsar's moment of inertia $I=10^{45}$~g~cm$^{2}$ and a $90^{\circ}$
  offset between its rotational and magnetic axes.}
\end{deluxetable}

We zapped RFI interactively with {\tt pazi} and used standard profiles to generate four TOAs per epoch with {\tt pat} -- summing across time and averaging down to four frequency subbands. Most of our observations were taken at 350~MHz, so retaining some frequency-dependence in our TOAs allowed us to fit for a linear slope in DM over our entire data span ($d\,{\rm DM}/dt$). In order to phase-connect the entire dataset, we fit for spin, position, proper motion, DM, and binary parameters (see Table \ref{tab:1400par}). Parameter fits were carried out with {\sc Tempo}\footnote{http://tempo.sourceforge.net/} timing software and the DE421 Solar System ephemeris; the timing solution is referenced to UTC (NIST). Due to \msp's small eccentricity, we used the ELL1 binary model, described originally in Appendix A of \cite{lcw+01}. The parameter uncertainties shown in Table \ref{tab:1400par} reflect 1-$\sigma$ (68\%) uncertainties on measured parameters. However, a global, multiplicative error factor (EFAC) has been applied to individual TOA errors such that the resulting reduced $\chi^2$ value is one. Fitting for all parameters in our current timing solution results in 4 \us root-mean-square (RMS) residuals with no obvious systematic trends (see Figure \ref{fig:res}). 

The position reported in Table \ref{tab:1400par} differs from that published in \cite{rsm+13} by $6.7\arcmin$; that timing solution was based on data spanning less than a year and therefore, was likely affected by position/spin-down covariance. We found an initial phase-coherent timing solution for \msp spanning several years in late June of 2015 and started observing it using the corrected position shortly afterwards (MJD 57199). For 350/820~MHz GBT observations, a $6.7\arcmin$ position offset results in a 9/43\% degradation in gain respectively. Since the majority of our timing observations were conducted at 350~MHz, the offset did not result in a significant loss of sensitivity.

\begin{figure}
    \centering
    \includegraphics[width=0.45\textwidth]{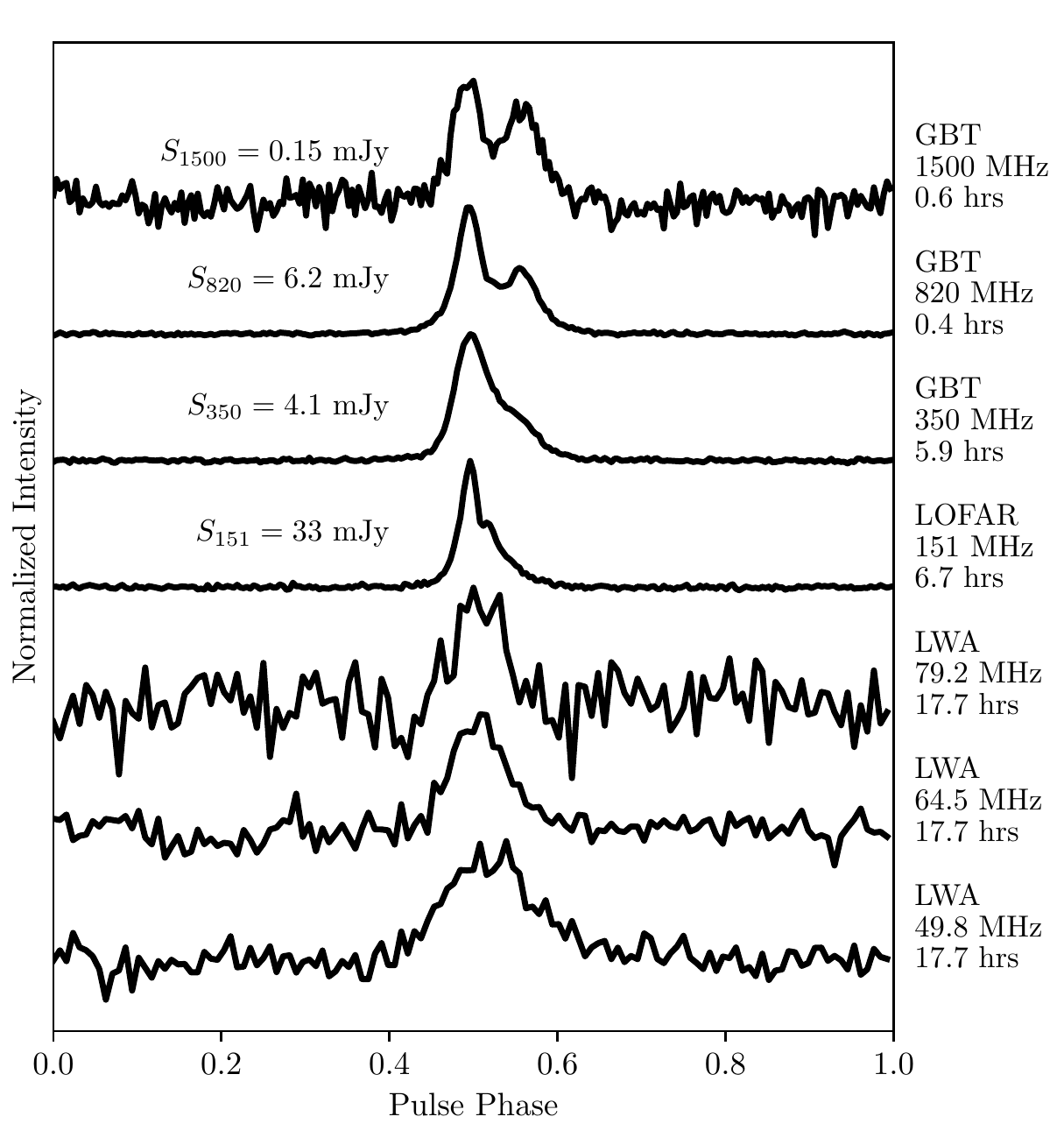}
    \caption{Integrated profiles for \msp show radio intensity as a function of pulse phase at a variety of observing frequencies spanning 50$-$1500~MHz. Profiles obtained with LWA and LOFAR/GBT observations are plotted here with 128 and 256 bins, respectively. Frequency-dependent flux values (e.g. S$_{\rm 1500}$) are shown next to corresponding profiles, each with $\sim50\%$ uncertainty.}
    \label{fig:profile}
\end{figure}

\subsection{Flux Density Estimates \& Scintillation}

We re-folded existing data and aligned profiles using our new timing solution, then summed profiles from separate frequency bands in-phase using {\tt psradd} (see Figure \ref{fig:profile}). Figure \ref{fig:profile} also includes relatively short test scans taken with the GBT at 820~MHz and 1500~MHz at the best-fit timing position. With GBT data, we estimated flux densities between 350$-$1500~MHz by measuring signal-to-noise ratios in each case and applying the radiometer equation \citep[see e.g.,][]{lk+04}. \psr\ was first detected at low frequency in a LOw-Frequency ARray \citep[LOFAR;][]{vwg+13} census of MSPs \citep{kvh+16}, but we obtained additional data for further study to generate the profile shown in Figure \ref{fig:profile}. With LOFAR data, we measured calibrated flux densities from 15 observations conducted over a $\sim$6 month period and quote the median value with 50\% uncertainties ($S_{151}=33\pm16$~mJy) since we did not carefully account for flux density variations due to \msp getting close to the Sun during this observing campaign and difficulties in calibrating LOFAR pulsar flux density measurements \citep{mkb+17}. We assume similar uncertainties for GBT flux density estimates, although they are likely even higher for nominal $S_{820}$ and $S_{1500}$ values since we do not yet have enough detections in these bands to average over flux density variability due to scintillation and other effects. Although scintillation may still be problematic for consistent detectability given its low DM, test observations at 820~MHz and 1500~MHz suggest that \msp should be re-evaluated for PTA inclusion. 

Finally, Figure \ref{fig:profile} shows summed profiles for \msp in three frequency bands (49.8~MHz, 64.5~MHz, and 79.2~MHz \--- each with 19.6~MHz bandwidth) obtained with the Long Wavelength Array \citep[LWA; e.g.][]{tek+12}. As of 2015, only three other MSPs were detected in an initial census \citep{srb+15}, so \msp is one of very few MSPs detected at these low frequencies. Since we have not yet carefully accounted for flux density variations due to a variety of known factors (e.g. frequency, zenith angle, and local sidereal time), we omit flux density estimates for the LWA detections shown in Figure \ref{fig:profile}.

Due to \msp's low DM, we expect it so scintillate heavily, and we see evidence of this in the significantly tailed distribution of 350~MHz TOA weights. However, looking at dynamic spectra from individual observations, there are no visible scintles, indicating that the scintillation timescale and bandwidth are too large to be resolvable by these observations. Because the scintillation timescales and bandwidths are not measurable in our data, we rely on estimates from the NE2001 \citep{cl+02} electron density model to better understand \msp's scintillation behavior. For the GBT profiles shown in Figure 2, only the one at 350~MHz incorporates enough data to average out the effect of scintillation. That is, the total integration time (5.9~hours) far exceeds the scintillation timescale at 350~MHz ($\Delta t_{\rm DISS, 350}\approx25$~mins). In all cases, scintillation bandwidths are comparable to our observing bandwidths, but at higher frequencies, the scintillation timescales ($\Delta t_{\rm DISS, 820}\approx35$~mins and $\Delta t_{\rm DISS, 1500}\approx45$~mins) exceed the total integration time for each profile. This suggests that corresponding estimated flux densities in these cases do not properly account for the effects of scintillation and are therefore somewhat biased.

\subsection{Constraining Distance}\label{sec:dist_constraints}
\begin{figure}
    \centering
    \includegraphics[width=0.48\textwidth]{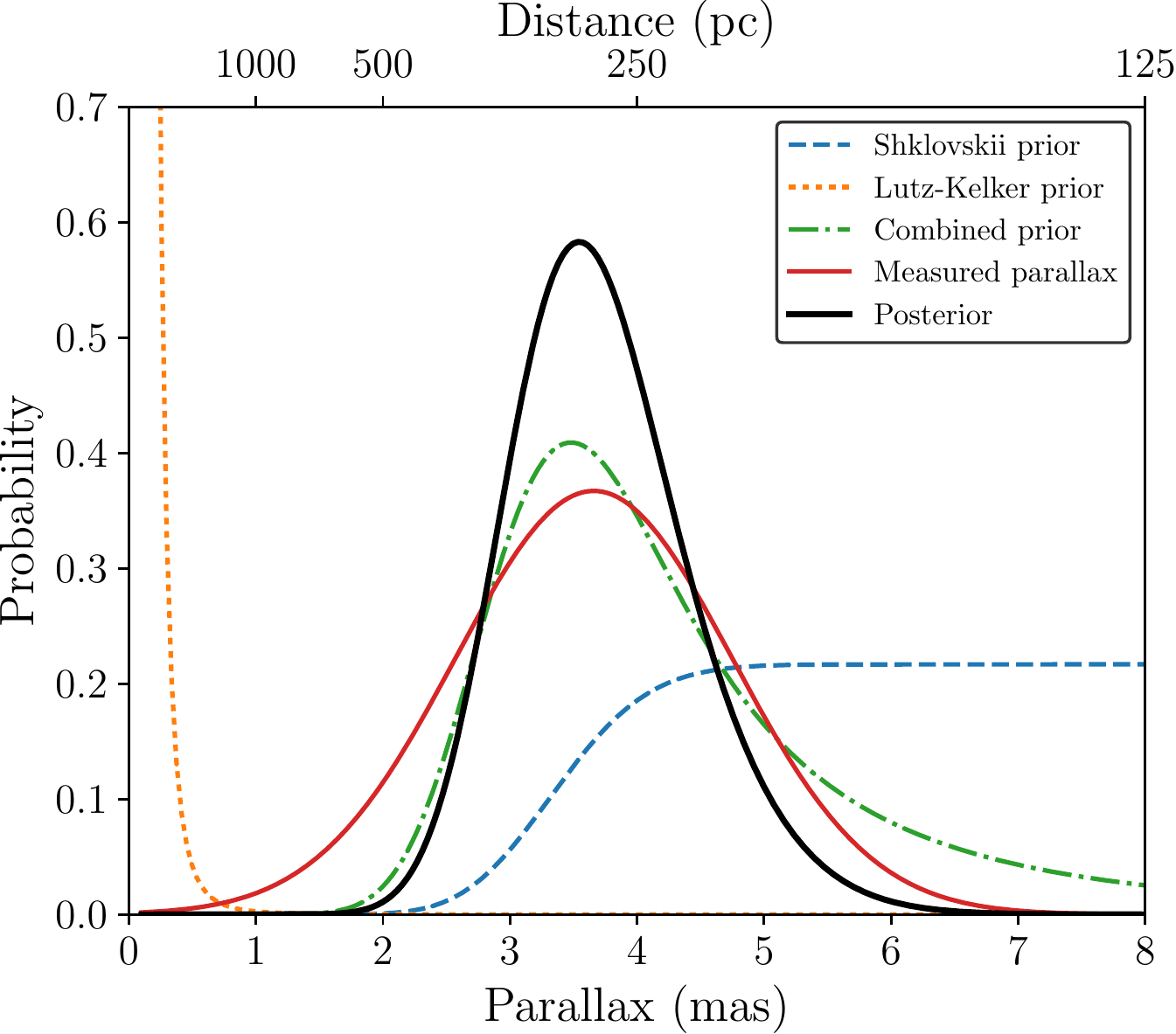}
    \caption{Posterior probability distribution function (black line) for the
      distance of \psr, based on Eqn.~\ref{eqn:varpi} for an intrinsic
      spin-down $\dot f=0$.  We also show the distribution from the
      measured parallax (red line), the prior derived from the
      limit on the distance due to the Shklovskii effect (blue dashed
      line), the volumetric prior for the Lutz-Kelker correction
      (Eqn.~\ref{eqn:LK}; orange dotted line), and the combined prior
      distribution (green dash-dotted line).
    \label{fig:distance}}
\end{figure}

Given \msp's position and dispersion measure (${\rm DM}=4.9$~\dmu), Galactic electron density models provide distance estimates along the pulsar's line of sight: 270~pc \citep{tc+93}, 500~pc \citep{cl+02}, and most recently, 350~pc \citep{ymw+17}. Normally, DM distances can be highly uncertain, particularly for pulsars with high Galactic latitudes like \msp ($b=45^{\circ}$). In comparison with earlier Galactic electron density models, \cite{ymw+17} improve on distance estimates for pulsars with $|b|>40^{\circ}$ whose distances have been measured independently. For 80\% of these pulsars, \cite{ymw+17} predict DM distances with uncertainties $<40\%$, but for some nearby MSPs we can measure distances to higher precision with pulsar timing. In some cases, the curvature of incoming wavefronts \citep{bh+86} can be measured as a 6-month periodic signature in timing residuals with amplitude,
\begin{equation}\label{eq:px}
A_\varpi = \frac{l^2\cos^2\beta}{2\,c\,d},
\end{equation}
where $A_\varpi$ is the amplitude of the timing parallax signature, $\beta$ and $d$ are the pulsar's ecliptic latitude and distance respectively, $l$ is the Earth-Sun distance (1 AU) and $c$ is the speed of light. Because of nearby distance estimates and its low ecliptic latitude ($\beta=2.1^{\circ}$), we decided to include parallax in \msp's pulsar timing model (see Table \ref{tab:1400par}) and detected it ($\varpi=3.6\pm1.1$~mas) with $\sim$3-$\sigma$ significance. This measurement constrains the system's distance inside the range $170<d/{\rm pc}<710$ with 95\% confidence, but the distribution is weighted towards larger distances since $d\propto1/\varpi$ (see cyan curve in Figure \ref{fig:distance}). We further refined these distance constraints using additional astrometric information.

Originally shown by \cite{shklovskii+70}, the induced period derivative  due to secular acceleration ($\dot{P}_{\rm Shklov}$) can account for a significant fraction of the measured spin-down ($\dot{P}_{\rm meas}$), which is composed of both intrinsic and kinematic components, $\dot{P}_{\rm meas}=\dot{P}_{\rm int}+\dot{P}_{\rm Shklov}$. Following \cite{nt+95}, we also investigated the contributions on $\dot{P}_{\rm meas}$ due to the pulsar's acceleration perpendicular to the Galactic plane ($2.4\times10^{-22}$~s/s, or 3\% of $\dot{P}_{\rm meas}$) and due to differential Galactic rotation ($1.4\times10^{-23}$~s/s, or 0.2\% of $\dot{P}_{\rm meas}$). These effects are more than an order of magnitude smaller than $\dot{P}_{\rm Shklov}$, so we consider them negligible for the discussion that follows. Assuming \msp is spinning down ($\dot{P}_{\rm int}>0$) and by imposing the constraint $\dot{P}_{\rm meas}>\dot{P}_{\rm Shklov}$, we place an upper limit on the pulsar's distance and therefore, a lower limit on its parallax. 

\begin{figure}[ht!]
\begin{center}
\includegraphics[width=0.50\textwidth,angle=0]{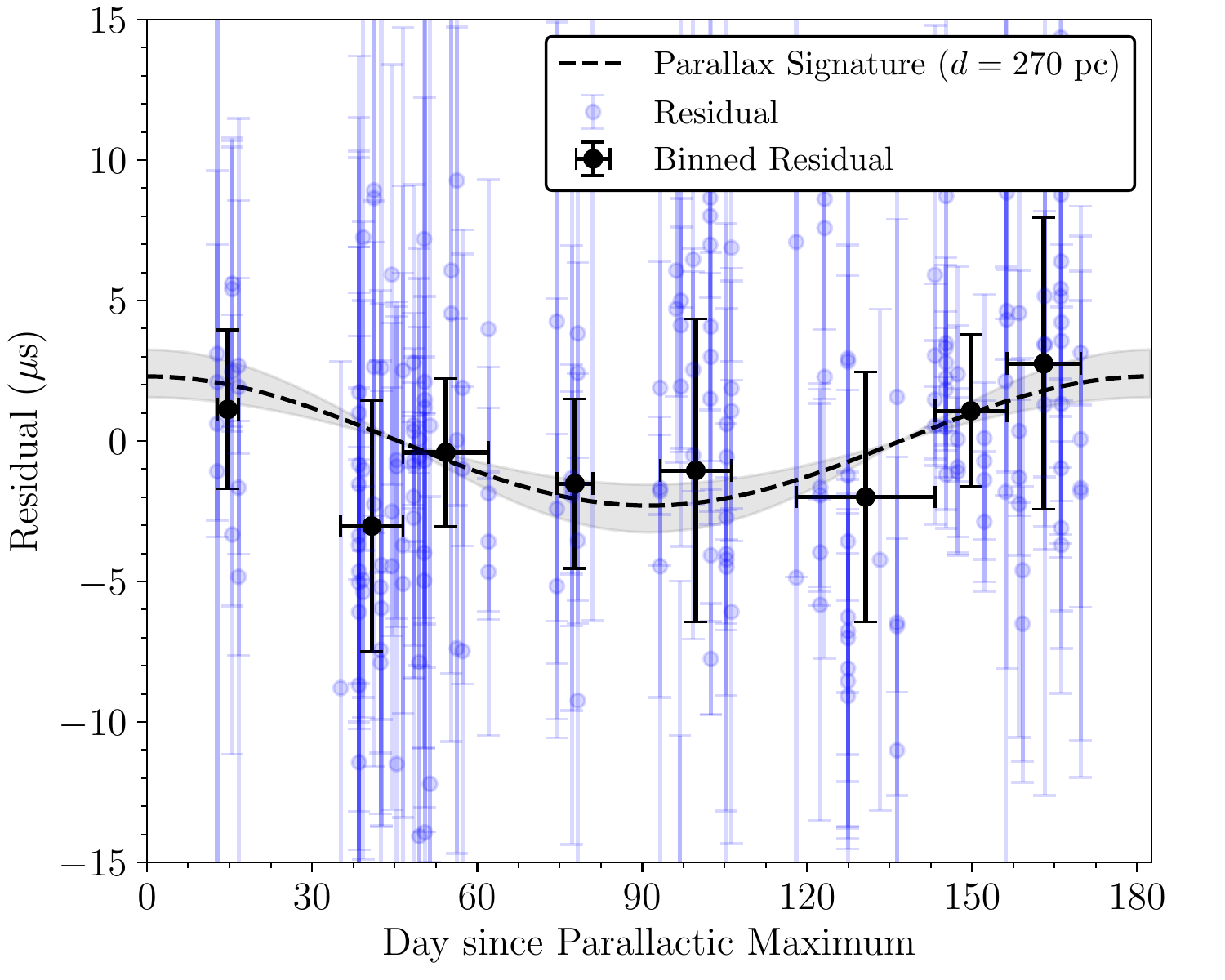}
\caption{Timing residuals and uncertainties for \msp\ (light blue), folded over a 6-month period and binned over stretches of less than 20~days. Day zero is defined as the time at which the Earth-Sun-pulsar angle is 90$^{\circ}$ (parallactic maximum). Black points show residuals' averages, weighted by their uncertainties squared. Due to uneven sampling, groups of residuals spanning $\leq20$~days were chosen to exclude gaps longer than 10 days and horizontal error bars show the extent of residuals that were averaged. The shaded region shows the range of expected parallax signature amplitudes in the residuals for distances that correspond to limits from our 95\% confidence interval, $190<d<400$~pc ($6.5>A_\varpi>3.1$~\us). The dashed line shows $A_\varpi=4.6$~\us, the expected amplitude of a timing parallax signature corresponding to the system's highest probability distance, $d=270$~pc.}
\label{fig:res2}
\end{center}
\end{figure}

We constrain the distance jointly through the parallax measurement and
the Shklovskii effect, also applying corrections for the Lutz-Kelker
bias \citep{lk+73}.  Adopting the notation of \citet{vlm10},
we attempt to determine the true parallax $\varpi$ given the
measurement $\varpi_0$ via,
\begin{equation}
p(\varpi|\varpi_0) = \frac{p(\varpi_0 |
  \varpi)p(\varpi)}{p(\varpi_0)},
\end{equation}
where we use a normal distribution for $p(\varpi_0|\varpi)={\cal
  N}(\varpi_0,\sigma_\varpi)=\exp(-(\varpi_0-\varpi)^2/2\sigma_\varpi^2)/\sqrt{2\pi\,\sigma_\varpi^2}$ and take $p(\varpi_0)$ to be flat.
We use a volumetric prior for $\varpi$ to account for the Lutz-Kelker
bias,
\begin{equation}
p_D(\varpi) \propto \varpi^{-4},
\label{eqn:LK}
\end{equation}
and add an additional term to the prior to account for the Shklovskii
effect.  We infer a distribution on the distance based on the 
proper motion $\mu$ and spin-down,
\begin{equation}
\varpi_{\rm Shklov} = \bigg(\frac{-f}{c~(\dot f_{\rm meas}-\dot f_{\rm int})}\bigg)~\mu^2 = A\,\mu^2,
\end{equation}
with $A=-f/c~(\dot f_{\rm meas}-\dot f_{\rm int})$.
We take the proper motion to be given by $p(\mu_0|\mu)={\cal
  N}(\mu_0,\sigma_\mu)$. Note that we have implicitly assumed that the parallax and proper motion distributions are independent (i.e. not correlated), but have verified this through exploration of the parameter space and believe it to be a robust assumption. Then, with the constraint that $\dot f_{\rm int}\leq 0$, we get a lower limit on $\varpi$ given by the cumulative integral of the distribution of $p(\mu_0|\mu)$ transformed to $\varpi$,
\begin{equation}
p_\mu(\varpi) = \int_0^\varpi d\varpi^\prime \frac{1}{\sqrt{8\pi A\, \varpi^\prime\,\sigma_\mu^2}}~e^{-(\sqrt{\varpi^\prime/A}-\mu_0)^2/2\sigma_\mu^2}
\label{eqn:varpi}
\end{equation}
suitably normalized.   Our final prior distribution $p(\varpi)$ is the product of
$p_D(\varpi)$ and $p_\mu(\varpi)$, resulting in 95\% confidence intervals on 
parallax and distance of $\varpi=3.7^{+1.6}_{-1.2}$~mas and $d=270^{+130}_{-80}$~pc, respectively. 
We use the confidence interval on distance to show corresponding parallax signatures  
in Figure \ref{fig:res2}, computed using Equation \ref{eq:px}. In this figure, we also show 
binned timing residuals to illustrate the
parallax signature measured with pulsar timing techniques described earlier.

We checked the parallax fit with a bootstrap method \citep{efron+1979}, generating 50,000 sets of TOAs by randomly sampling the original TOAs with replacement until each trial set had the same number of TOAs as the original. Starting with our best-fit timing solution, we re-fit for all parameters using each trial TOA file and recorded trial fit parameters.

Overall, the bootstrap reproduced the conclusions from our best-fit timing solution once we excluded non-physical results (such as negative parallax). The widths of the bootstrap posterior distributions for individual parameters were somewhat larger than uncertainties reported by {\sc Tempo}, by a factor of 1--2 depending on the parameter. However, our conclusions remain largely unchanged: even if we assume a factor of 2 increase in the parallax uncertainties, the effect on the 95\% confidence interval for the distance is negligible, going from 190--400~pc to 160--420~pc. We are obtaining more data as well as investigating further timing techniques to fully reconcile this issue.

\begin{figure*}[h]
    \centering
    \includegraphics[width=0.9\textwidth]{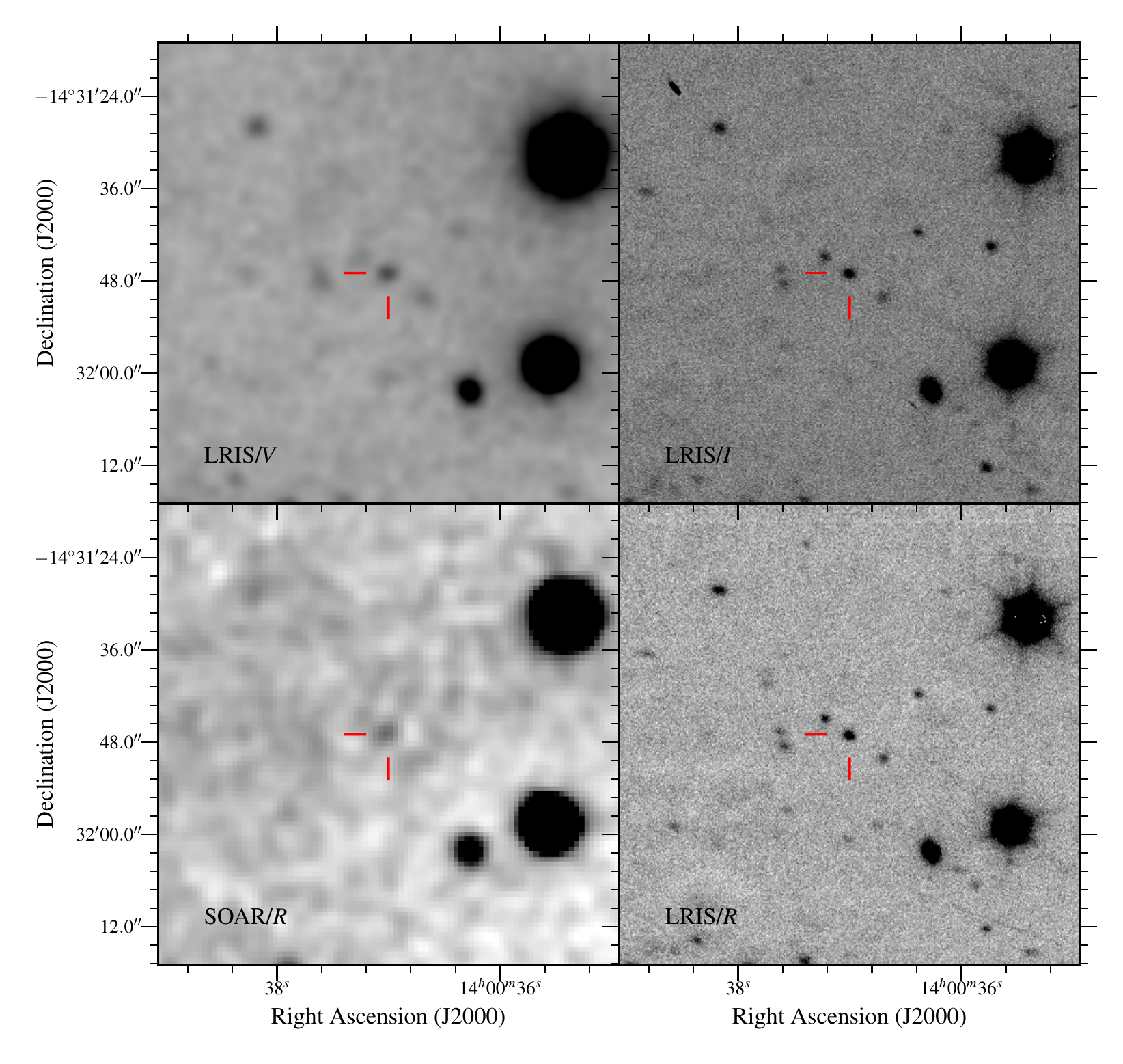}
    \caption{The region around the radio position of \psr, with data
      from Keck I/LRIS in the $V$ (upper left), $R$ (lower right) and
      $I$ (upper right) filters.  We also show the same region with
      data from SOAR/Goodman in the \Rc\ filter.  Each image is
      $1\arcmin$ on a side, with north up and east to the left.  The
      radio position of \psr\ is shown with the tick marks.  For
      the LRIS/$V$ and SOAR/\Rc\ images we have additionally smoothed
      the data to improve the visibility of the counterpart.}
    \label{fig:detects}
\end{figure*}

\begin{figure*}
  \plottwo{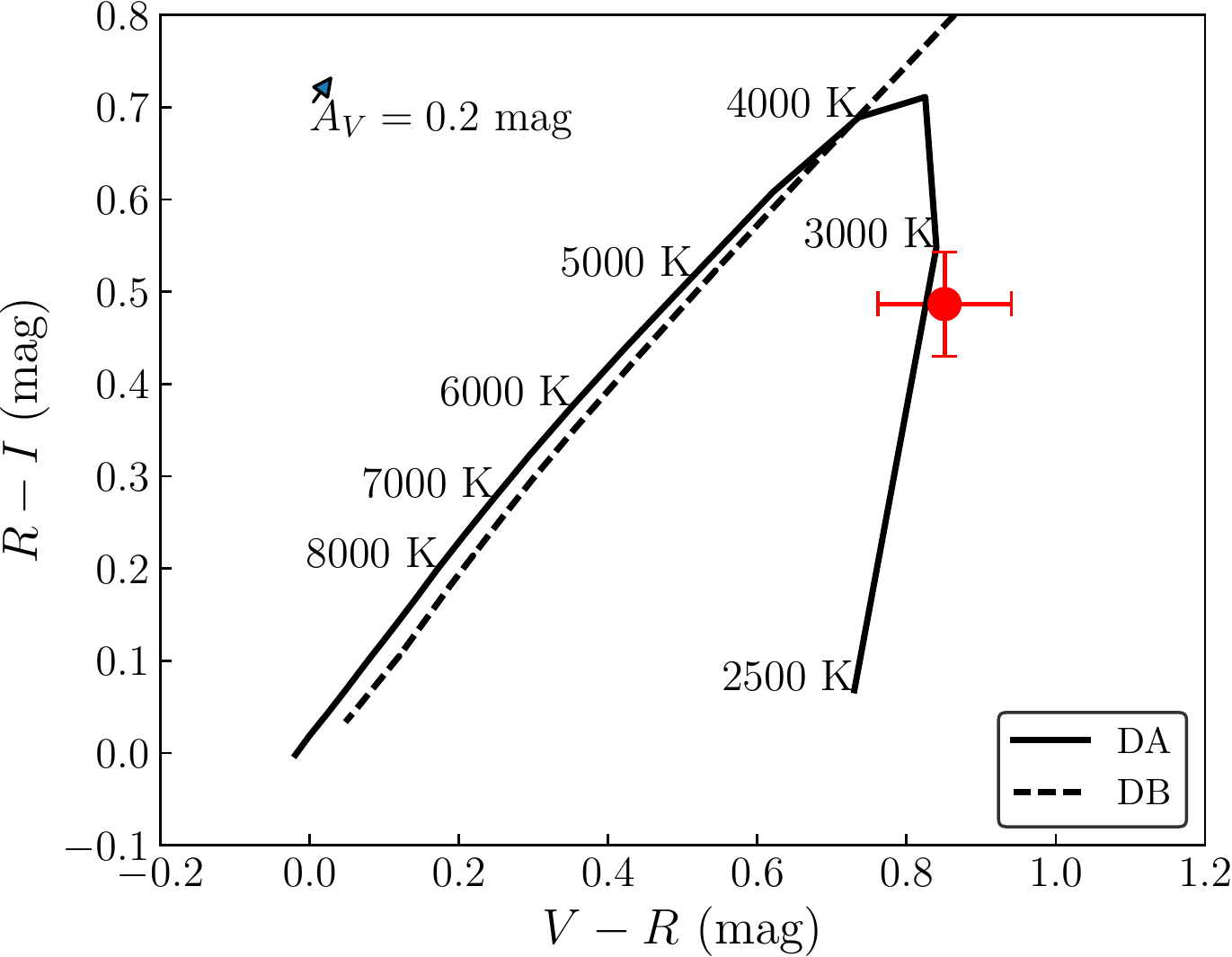}{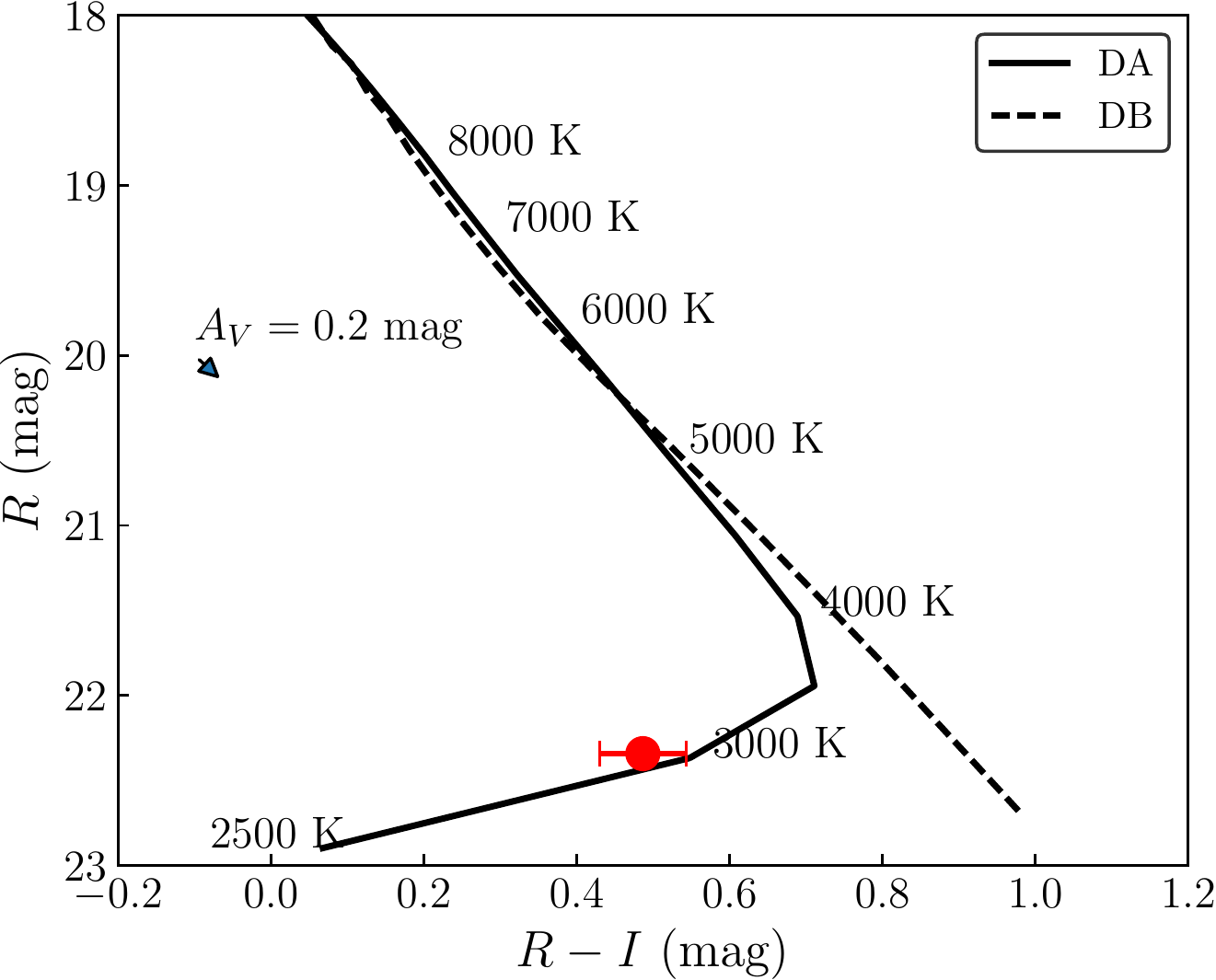}
  \caption{Color-color (left) and color-magnitude (right) diagrams for
    \psr, based on the photometry in Table~\ref{tab:optical}.  The
    color-color diagram shows the $R-I$ color vs.\ the $V-R$ color
    along with synthetic photometry from  \citet{tbg+11} and
    \citet{bwd+11} for hydrogen (DA; solid line) and helium (DB;
    dashed line), respectively.  The synthetic photometry is labeled
    with the effective temperature, and the arrow shows a reddening vector
    for $A_V=0.2$.  The color-magnitude diagram shows the $R$
    magnitude vs.\ the $R-I$ color with the same synthetic photometry
    models, which have been adjusted to have a radius of
    0.0219~\Rsun at a distance of 270~pc.}
  \label{fig:color}
\end{figure*}

\section{Optical Follow-up}\label{sec:optical}
We used the Goodman Spectrograph on the 4.1-m SOAR Telescope \citep{cca+04} in its imaging mode to obtain optical photometry of a $6'\times6'$ field surrounding \psr.
The object frames were bias-subtracted and flat-fielded using {\tt CCDPROC} and other standard routines in IRAF\footnote{IRAF is distributed by the National Optical Astronomy Observatories, which are operated by the Association of Universities for Research in Astronomy, Inc., under cooperative agreement with the National Science Foundation.} \citep{tody+86} and averaged together using the {\tt IMCOMBINE} routine to create a final, master frame.  We then ran the master frame through {\it astrometry.net} \citep{lhm+10} to obtain an astrometric calibration to a precision of better than 0.1$\arcsec$.

A visual inspection of the master object frame, a subset of which is shown in Figure \ref{fig:detects}, reveals a faint optical source at the precise location of \msp determined from the radio observations. We used the {\tt PHOT} task in the IRAF/DAOPHOT package to extract aperture photometry of nearly two dozen stars in the field of view of the master frame, covering a range of magnitudes $R_{\rm c} \simeq 15-20$.  Our measured magnitude for each star was compared to the values reported by \citet{qyb+15} in order to determine the zero--point magnitude of our data set, after converting their $R_F$ photographic red band magnitudes to $R_{\rm c}$ via the transformations of \citet{bessel+86}.  We then used {\tt PHOT} to perform aperture photometry on the optical component of \msp and derived a final $R_{\rm c}$-band magnitude of $R_{\rm c} = 22.5 \pm 0.3$.

We obtained additional, deeper imaging of \msp\ using the blue and
red sides of the Low-Resolution Imaging Spectrometer (LRIS;
\citealt{occ+95}) on the 10-m Keck~I telescope.  The data were reduced
using standard procedures in IRAF, subtracting the bias,
dividing by flatfields, and combining the individual exposures. At
this time \msp\ was only visible during the very beginning of the
night, so the observations were obtained at somewhat high airmass
(up to 2.0).

Guided by the SOAR detection, we were able to detect the counterpart
to \msp\ in all three bands of the LRIS imaging  as seen in
Figure~\ref{fig:detects}.  We reduced the LRIS data using standard procedures 
provided by the {\sc LPipe} reduction framework.\footnote{http://www.astro.caltech.edu/$\sim$dperley/programs/lpipe.html}  
Astrometric calibration was performed against USNO-B \citep{mlc+03} and 
RMS scatter against the catalog was $\sim$0.4$''$ for 16-19 matched sources.
Aperture photometry was measured using {\tt SExtractor} \citep{ba+96}. 
We photometrically calibrated the LRIS
images using the Pan-STARRS $3\pi$ Steradian Survey
\citep{cmm+16,fmc+16} catalog.\footnote{The counterpart is visible directly in Pan-STARRS (PS1) stacked $r$ and $i$ band images, but is not listed in the corresponding catalog, suggesting a low significance detection.  In any case we did not use PS1 to motivate followup because the data were released after discovery of the counterpart with SOAR.} In each image we identified $\sim$20
stars that matched those from the catalog and were additionally not
extended, saturated, or otherwise affected by bad pixels.  We
transformed the Pan-STARRS photometry to the Johnson-Cousins system
using the results from \citet{tsl+12} and determined zero-points for
each LRIS image. Comparing observations of 15 other stars detected by
both SOAR (in \Rc) and Keck (in $R$), we found consistent results.

In Figure~\ref{fig:color} we plot these results on color-color and
color-magnitude diagrams along with the predictions of model atmospheres for hydrogen
(DA) and helium (DB) white dwarf atmospheres from \citet{tbg+11} and \citet{bwd+11}, 
respectively.\footnote{\url{http://www.astro.umontreal.ca/$\sim$bergeron/CoolingModels/}}
From the color-color diagram it appears that the $R-I$ color is
consistent either with an effective temperature $\Teff\approx 4800\,$K
or $\Teff\approx 3000\,$K. This degeneracy is a result of
collisionally-induced absorption by molecular H$_2$
\citep*{bsw+95,hansen+98}, which shifts flux from the near-infrared into the
optical.  However, from the $V-R$ color only the cooler solution seems
plausible.  Fitting the extinction-corrected photometry as a function
of $\Teff$ and angular size, we get a good solution for
$\Teff=3000\pm100\,$K and $R/R_\odot=(2.19\pm0.03)\times10^{-2}\,(d/270~{\rm pc})$, 
where we have increased the
uncertainty on $\Teff$ to account for the coarseness of our atmosphere grid.

\begin{deluxetable*}{l c c c c l}
\tabletypesize{\small}
\tablewidth{0pt}
\tablecaption{\label{tab:optical}Summary of Optical Observations of \psr}
\tablehead{
\colhead{Telescope/Instrument} & \colhead{Date} 
& \colhead{Filter} & \colhead{Airmass} & \colhead{Exposure} &
 \colhead{Magnitude} \\
 & & & & \colhead{(sec)} &
}
\startdata
SOAR/Goodman & 2016-06-09 & \Rc\ & 1.81 & $46\times 5$ &  $22.5\pm0.3$\\
Keck I/LRIS(blue) & 2016-08-02 & $V$ & 2.03 & $180$ &  $23.41\pm0.08$\\
Keck I/LRIS(red) & 2016-08-02 & $R$ & 1.71 & $300$ &  $22.52\pm0.04$\\
Keck I/LRIS(red) & 2016-08-02 & $I$ & 1.92 & 300 &  $21.99\pm0.04$\\
\enddata
\end{deluxetable*}

\section{Gamma-ray Spectra and Timing} \label{sec:gamma}
The radio timing position of \psr reported in Table \ref{tab:1400par} is within $5.3'$ of the \fermi\ Large Area Telescope (LAT) source 3FGL J1400.5$-$1437 (which has a 95\% confidence error ellipse of size $7.0'\times4.7'$). A positional association was noted by \citet{3FGL} and in the following discussion, we analyze the $\gamma$-ray source to evaluate the likelihood of an association and to search for evidence of $\gamma$-ray pulsations.

For this analysis, we extracted Pass 8 data starting from 2008 August 4 (the beginning of LAT survey mode operation) and extending through 2017 March 1 (Mission Elapsed Time $239557517-510019205$).  We selected SOURCE class, front and back-converting events ({\tt evclass = 128} and {\tt evtype = 3}) combined during the intervals of good science data ({\tt DATA\_QUAL=1} and {\tt LAT\_CONFIG=1}) and restricted our events to those with a zenith angle less than 90$\degr$.  We selected events between 100~MeV and 100~GeV from a 15$\degr$ radius around the pulsar and performed a binned likelihood analysis over a $20\degr \times 20\degr$ region with 0.1$\degr$ pixels. Starting with a model based on the 3FGL catalog \citep{3FGL}, we modified the target source's spectral model to be an exponentially cutoff power law of the form,
\begin{equation}\label{eq:powerexpcut}
\frac{dN}{dE} = N_0 \left(\frac{E}{E_0}\right)^{-\Gamma} \exp\left(-\frac{E}{E_{\rm cut}}\right),
\end{equation}
with normalization $N_0$ in photons cm$^{-2}$ s$^{-1}$ MeV$^{-1}$, reference energy $E_0$, cutoff energy $E_{\rm cut}$, and photon index $\Gamma$. 
To perform the maximum likelihood fit, we used the {\tt P8R2\_SOURCE\_V6} instrument response functions with the \fermi\  Science Tools version v11r05p02 and the \texttt{NewMinuit} fitting function.\footnote{\url{https://fermi.gsfc.nasa.gov/ssc/data/analysis/documentation/}} The isotropic diffuse model was {\tt iso\_P8R2\_SOURCE\_V6\_v06.txt},\footnote{\url{http://fermi.gsfc.nasa.gov/ssc/data/access/lat/BackgroundModels.html}} with normalization left free, and the Galactic diffuse model \citep{aaa+16} was {\tt gll\_iem\_v06.fits}, with index and normalization left free. In the initial fit, we held all values at the 3FGL catalog values except for the spectral parameters for the target source, and the normalization for sources within 6$\degr$ of the target or flagged in the 3FGL catalog as being variable.  We inspected the residuals map and found that one additional source at $\alpha=218.281^{\circ}$, $\delta=-17.992^{\circ}$ was required to model the region, so this was added to the model. This source is positionally associated with the quasi-stellar object PKS 1430$-$178. The best-fit spectral parameters for the pulsar are presented in Table \ref{tab:lat}, where the ``Test Statistic'' (TS) is the source detection significance \citep{mbc+96}. The exponentially cutoff power law model is preferred to a pure power law with a confidence of 4-$\sigma$ (TS$_{\rm cut} = 2\Delta \log({\rm likelihood})$ between the model with and without the cutoff). We then used {\tt gtfindsrc} to get an improved localization for the LAT source, which gave a position of $\alpha=210.166^{\circ}$, $\delta=-14.535^{\circ}$ (only $0.7'$ from the radio timing position) with a 95\% confidence radius of $3.3'$.

\begin{figure}
    \centering
    \includegraphics[width=0.45\textwidth]{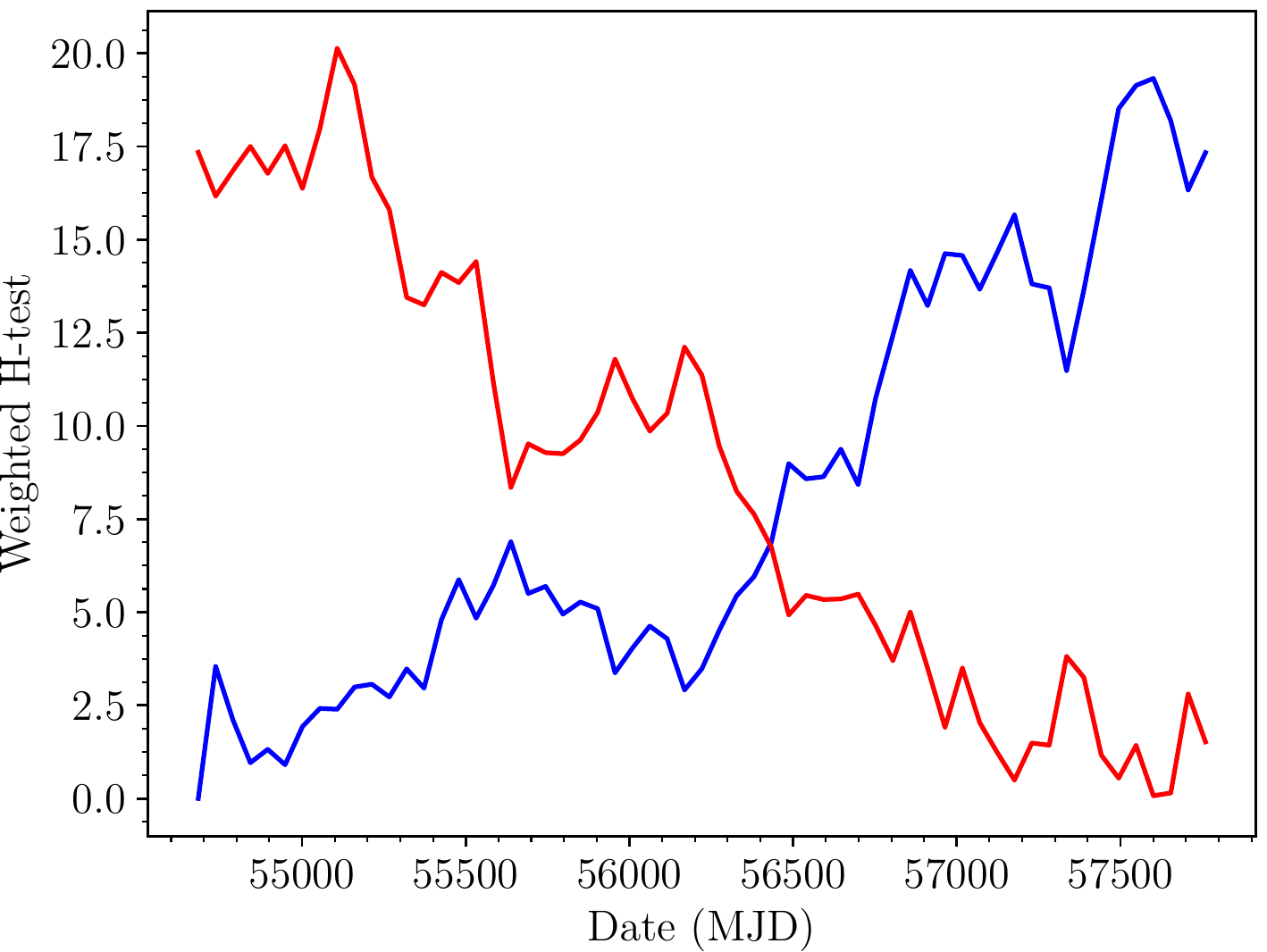}
    \caption{Weighted H-test vs time computed both forwards (blue) and backwards (red) in time. While the H-test does not reach the 5-$\sigma$ level, the rising H-test is indicative of a marginally-detected pulsation. The mostly monotonic rise is an indication that the pulse timing model used to fold the data is phase-coherent over the full LAT mission.
    \label{fig:htest}}
\end{figure}

For the timing analysis, we selected photons from a region of radius 2$\degr$ around the pulsar and assigned photon weights based on the best-fit spectral model. We computed a pulse phase for each selected LAT photon using the {\tt fermi} plugin for \textsc{Tempo2} \citep{rkp+11} and the best-fit radio timing model.  The pulsation significance was determined using the weighted H-test \citep{kerr+11} and the resulting H-test value was 17.4, corresponding to a significance of 3.3-$\sigma$ (see Figure \ref{fig:htest}). This is not sufficient to claim a secure detection, but suggests that weak LAT pulsations may be present from this source.

Although there is only weak evidence for the presence of pulsations, we find strong support for an association between 3FGL J1400.5$-$1437 and \psr, primarily due to their positional coincidence. The GeV spectrum of the 3FGL source shows significant curvature, providing additional support for an association. Also, $\Gamma$ and $E_\mathrm{cut}$ values are comparable to those of other MSPs in the Fermi Second Pulsar Catalog \citep[2PC;][]{2PC}.  Finally, the marginal detection of pulsations provides additional evidence in favor of the identification of the $\gamma$-ray source with the pulsar, though not with certainty. Assuming this association is real, we can compare it to the rest of the MSP population, which are often $\gamma$-ray emitters.

\begin{figure}
    \centering
    \includegraphics[width=0.45\textwidth]{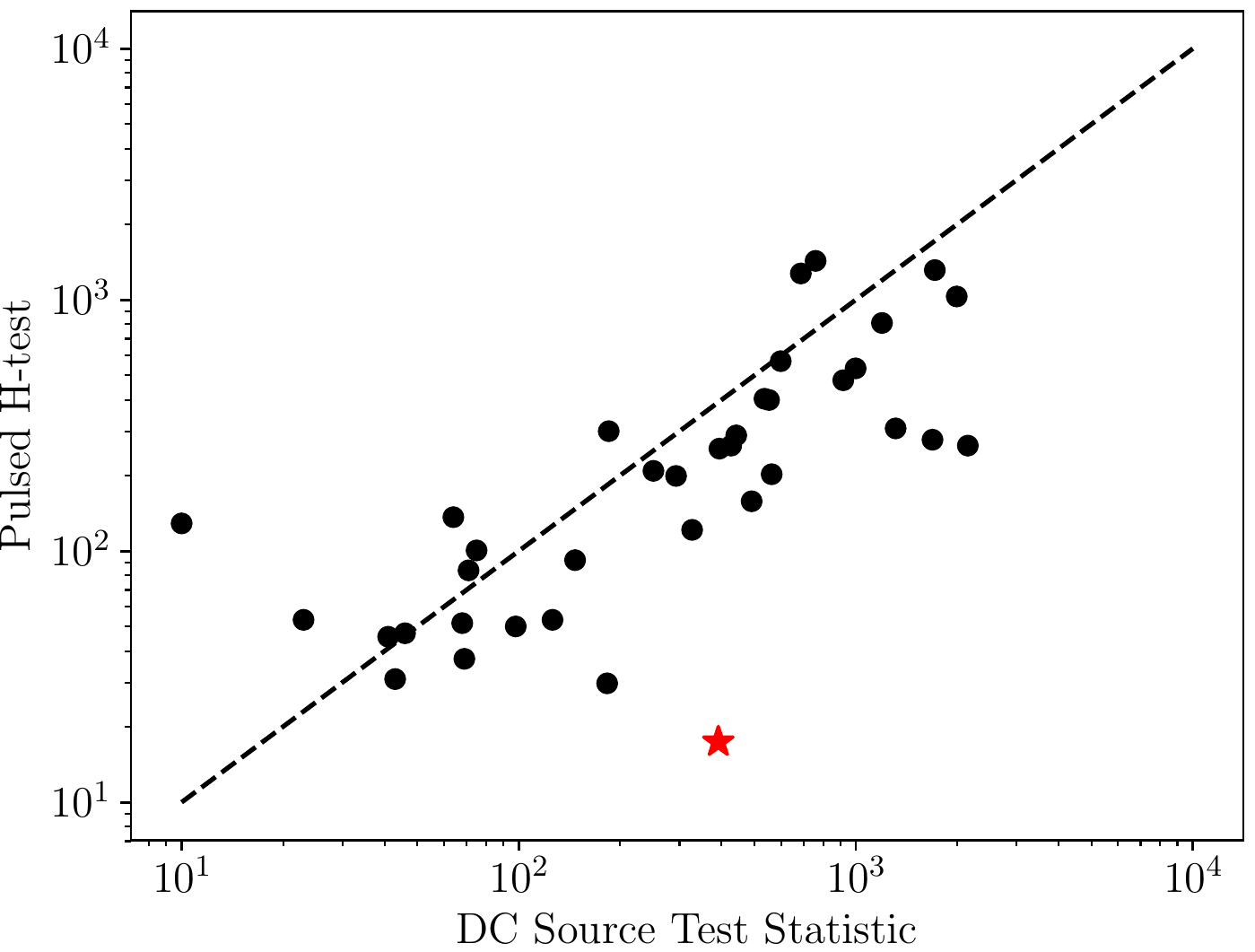}
    \caption{Weighted H-test statistic vs. Test Statistic for the DC $\gamma$-ray source for the sample of MSPs in 2PC \citep{2PC}. The red star shows \psr.
    \label{fig:hvsts}}
\end{figure}

Since the DC (constant/non-pulsed) $\gamma$-ray source is strongly detected (a TS of 391 corresponds to a detection significance of 17-$\sigma$) we might expect to see detectable pulsations.  Figure \ref{fig:hvsts} shows the correlation between DC source TS and the weighted H-test for pulsations, based on data from 2PC. Clearly, the pulsed significance for \msp\ is far below what is expected based on its $\gamma$-ray flux.  Assuming the pulse timing model is good, this indicates either a low pulsed fraction, or a sinusoidal (rather than sharply-peaked) pulse profile, or both, hampering the detection of pulsed emission.

\begin{deluxetable}{lllll}
\tabletypesize{\small}
\tablewidth{0pt}
\tablecaption{\label{tab:lat}LAT Spectral Analysis Results}
\tablecolumns{2}
\tablehead{\colhead{Parameter} & \colhead{Value}}
\startdata
3FGL Source\dotfill  & J1400.5$-$1437  \\
$\Gamma$\dotfill             & 2.1(1)         \\
$E_{\rm cut}$ (GeV)\dotfill         & 4.7(17)       \\
Photon flux\tablenotemark{a} (\fluxu)\dotfill & 20(2)          \\
Energy flux\tablenotemark{a} (\efluxu)\dotfill & 10.2(6)        \\
TS\dotfill                   & 391             \\
TS$_{\rm cut}$\dotfill    & 17.7 \\
\enddata
\tablenotetext{a}{Over the $0.1-100$~GeV energy range.}
\tablecomments{Quantities in parentheses are 68\% confidence uncertainties (statistical only) in the last digit.}
\end{deluxetable}

\section{X-ray Observations} \label{sec:xray}
\psr was targeted with the X-ray Multi-Mirror Mission, \textit{XMM-Newton} on 2016 July 17 for a duration of 39.8~ks (ObsID 0780670101; PI S.~Bogdanov). The European Photon Imaging Camera (EPIC) pn \citep{sbd+01} and MOS1/2 \citep{taa+01} instruments were configured in {\it full window mode} and used the thin optical blocking filters. We reprocessed the observation data files using the \emph{XMM-Newton} Science Analysis Software (SAS\footnote{The \textit{XMM-Newton} SAS is developed and maintained by the Science Operations Centre at the European Space Astronomy Centre and the Survey Science Centre at the University of Leicester.}) version {\tt xmmsas\_20160201\_1833-15.0.0}. The data were subjected to the standard flag, pattern, and pulse invariant filtering. Periods of strong background flares were excised, which resulted in effective exposures of 35.4, 36.3, and 28.2~ks for the MOS1, MOS2, and pn, respectively. The cleaned data sets were used for the X-ray spectroscopic analysis presented below. Due to the 0.73~s read-out time of the pn and 2.6~s for MOS1/2, it was not possible to fold the data at the MSP period to study any X-ray pulsations.

\begin{figure}
    \centering
    \includegraphics[scale=0.55]{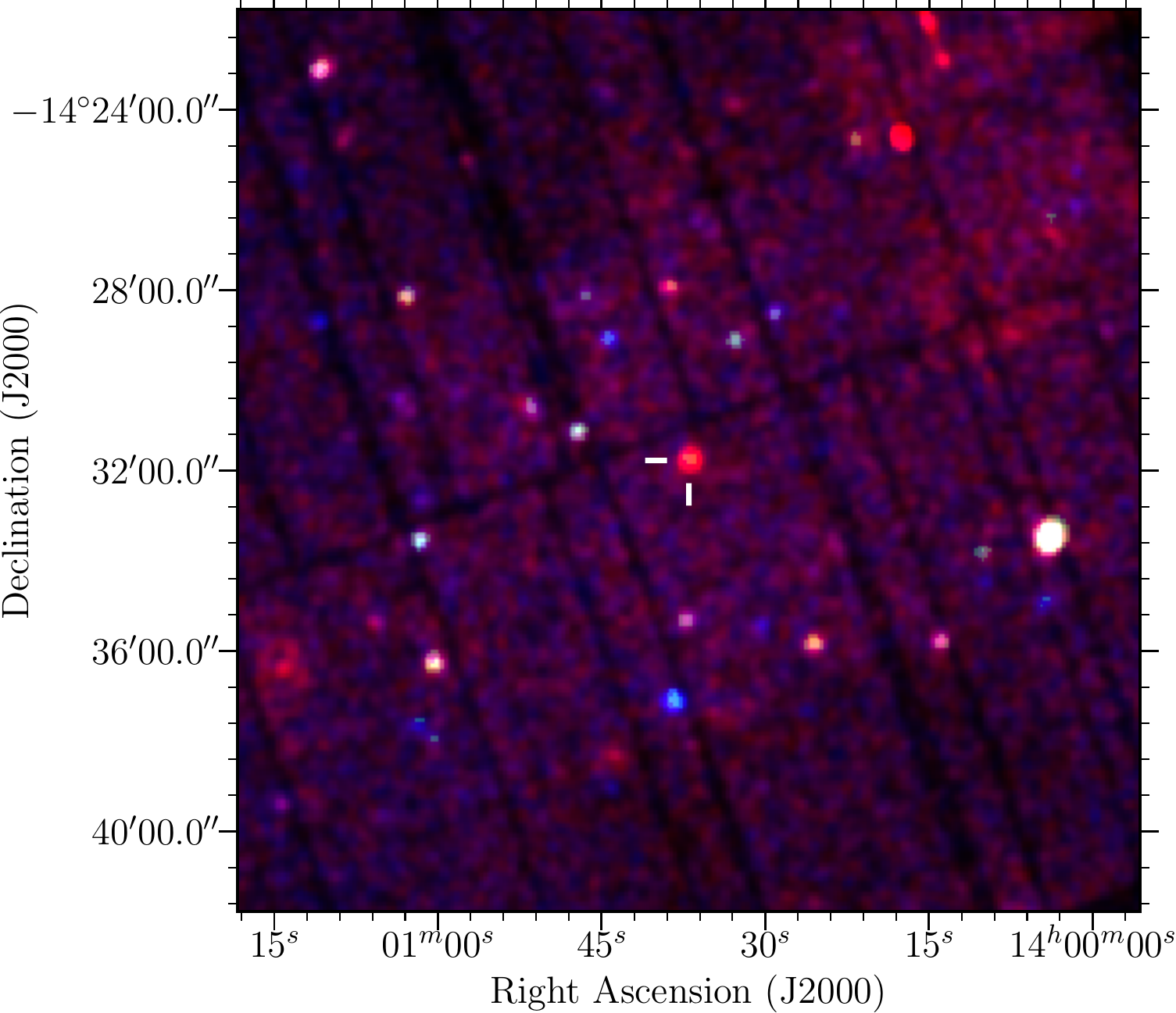}
    \caption{A ``true color" image of the combined \textit{XMM-Newton} EPIC MOS1, MOS2 and pn data of \psr with red corresponding to $0.3-1$~keV, green to $1-2$~keV, and blue to $2-7$~keV. The pulsar is a faint and soft X-ray source (typical of MSPs) and is marked by the two white ticks near the center of the image.}
    \label{fig:xmm_image}
\end{figure}

Figure~\ref{fig:xmm_image} shows the co-added representative color image from all three \textit{XMM-Newton} detectors. It is evident that \psr is a faint X-ray source and it is quite soft, with nearly all source photons detected below $\sim$1.5~keV. To produce spectra suitable for fitting, the pn, MOS1, and MOS2 data were grouped such that each energy bin contained at least 25 counts. The binned spectra from all three detectors were modeled jointly in XSPEC. Three single-component models were considered: a power law, a blackbody, and a non-magnetic neutron star hydrogen atmosphere model \citep[NSATMOS;][]{hrn+06}. Due to the limited photon statistics, in the spectroscopic analysis we fix the value of the equivalent atomic hydrogen column density, $N_{\rm H}=1.5\times10^{20}$~cm$^{-2}$, determined from the empirical relation between DM and $N_{\rm H}$ from \citet{hnk+13}. In all cases, the {\tt tbabs} model \citep{wam+00}  was used to account for the interstellar absorption along the line of sight. 

A fit with a power law produces statistically acceptable results ($\chi_{\nu}^2=1.02$ for 21 degrees of freedom) but requires an implausibly steep power law photon index ($\Gamma\approx6.5$). A blackbody model yields a temperature of $kT=0.15\pm0.02$~keV, an effective emitting radius of  $R_{\rm eff}= 0.06^{+0.05}_{-0.04}$~km, an unabsorbed flux of $(1.07\pm0.15)\times10^{-14}$~erg~cm$^{-2}$~s$^{-1}$ in the $0.3-10$~keV range, and $\chi_{\nu}^2=0.70$ for 21 degrees of freedom. Fitting a hydrogen atmosphere model assuming a neutron star with mass $1.4$~\Msun, radius 12~km, and distance 270~pc resulted in a best-fit with a redshift-corrected effective temperature $T_{\rm eff}=7.8^{+1.5}_{-1.3}\times10^5$~K, an emitting area that is $0.60^{+0.08}_{-0.04}$\% of the total neutron star surface area, an unabsorbed 0.3--10~keV flux of $(1.15\pm0.17)\times10^{-14}$~erg~cm$^{-2}$~s$^{-1}$, and $\chi_{\nu}^2=0.84$ for 21 degrees of freedom. The soft thermal spectrum of \psr is typical of the sample of MSPs detected in X-rays \citep{zavlin+06, bgh+06,fhc+14}. This thermal radiation likely originates from the magnetic polar caps of the pulsar, which are heated to $\sim10^6$~K by a return flow of relativistic particles from the open field region of the magnetosphere \citep[e.g.][]{hm+02}.



\begin{figure*}
    \centering
    \includegraphics[width=0.9\textwidth]{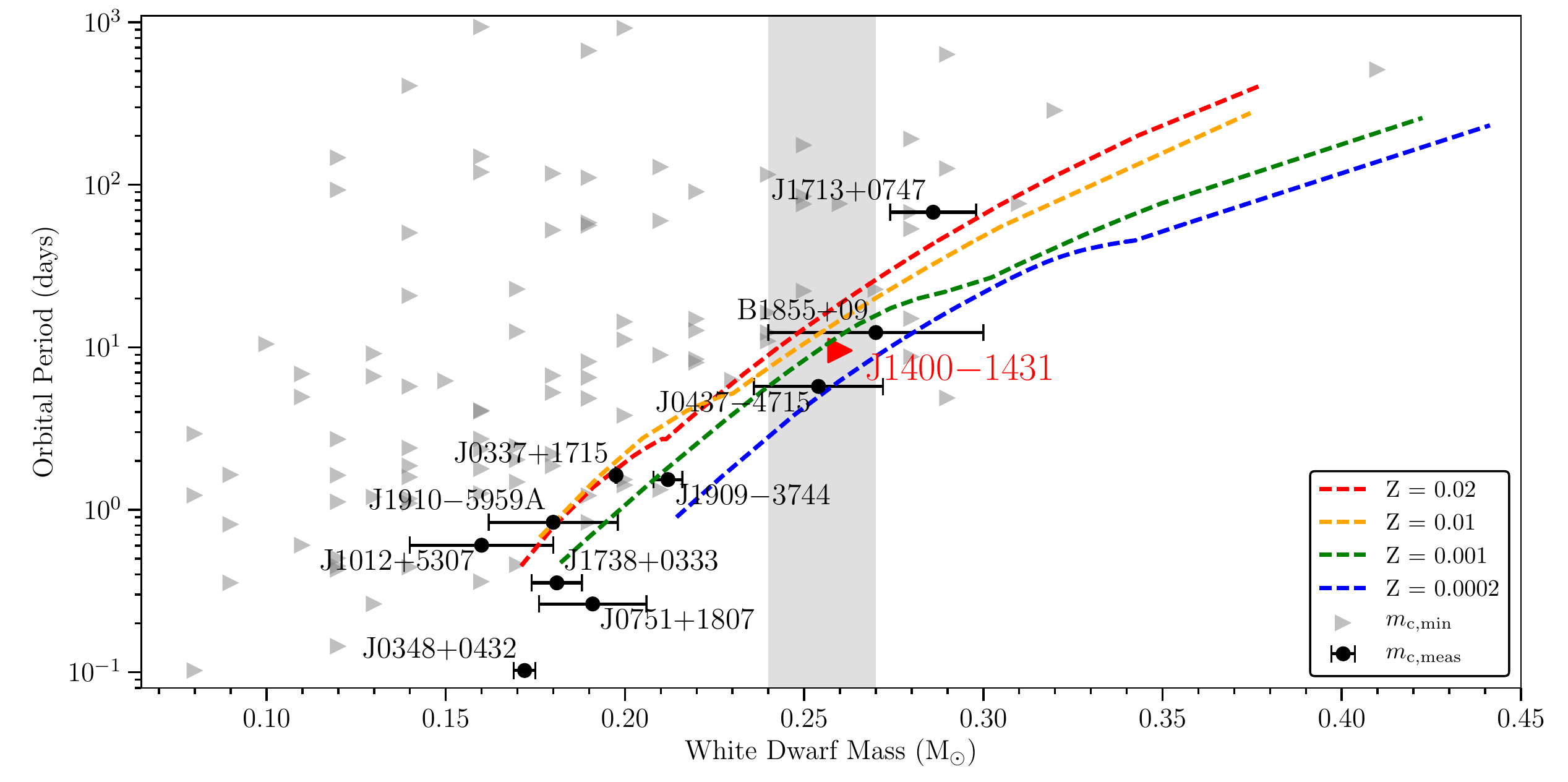}
    \caption{Colored dashed lines show the $(P_{\rm b},m_{\rm WD})$-relationship expected for He WD populations, simulated with corresponding metallicities listed in the legend \citep{imt+16}. Minimum companion masses (gray triangles) determined with pulsar timing and measured WD masses (black circles) from \cite{mht+05} are also plotted. The red triangle shows $m_{\rm c,min}=0.26$~\Msun derived with pulsar timing for \msp's WD companion and the gray shaded region indicates masses consistent with \cite{imt+16} models given $P_{\rm B}=9.5$~days, $0.24<m_{\rm c}<0.27$~\Msun.}
    \label{fig:PbMc}
\end{figure*}

\begin{figure*}
   \centering
    \plottwo{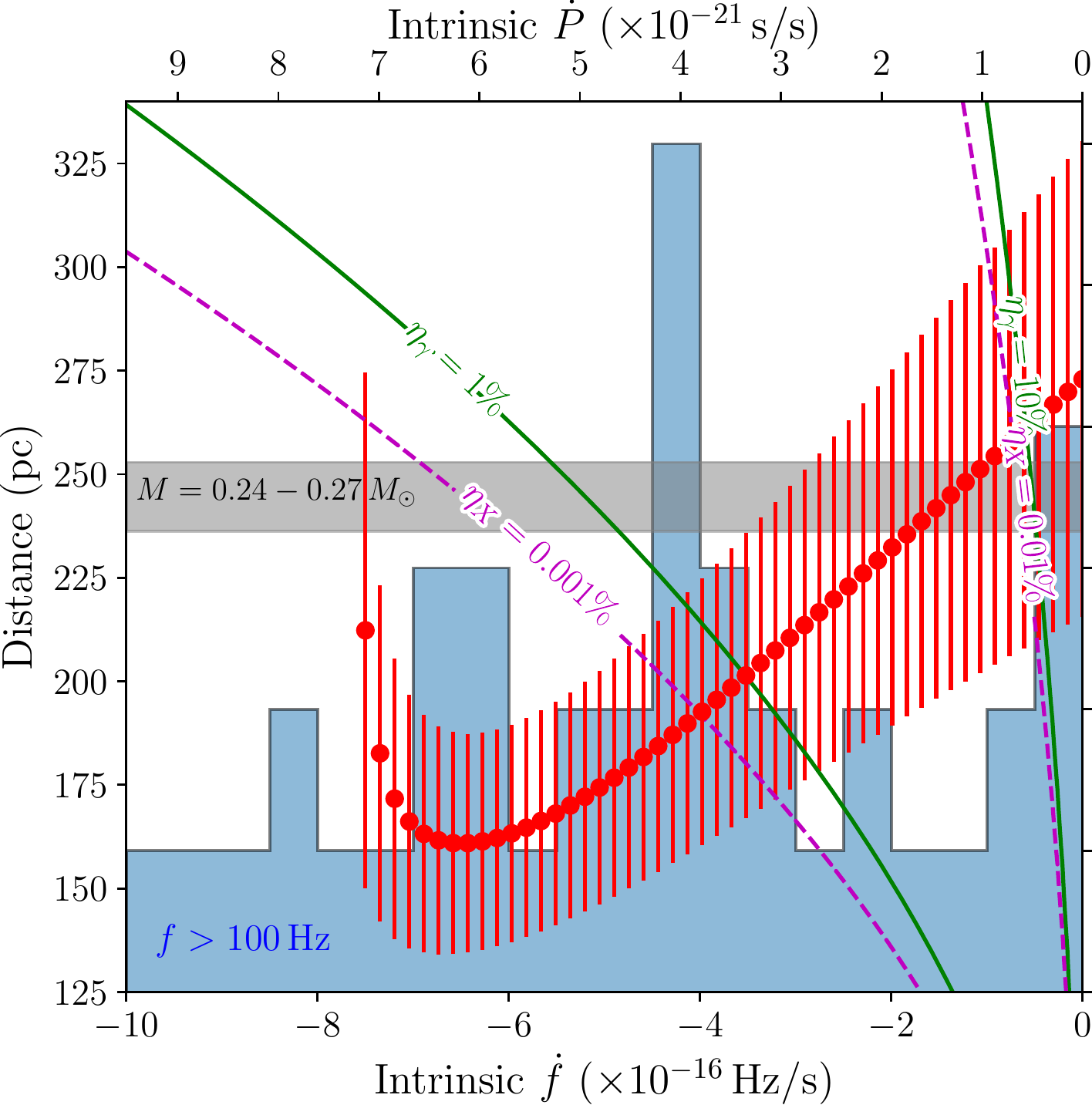}{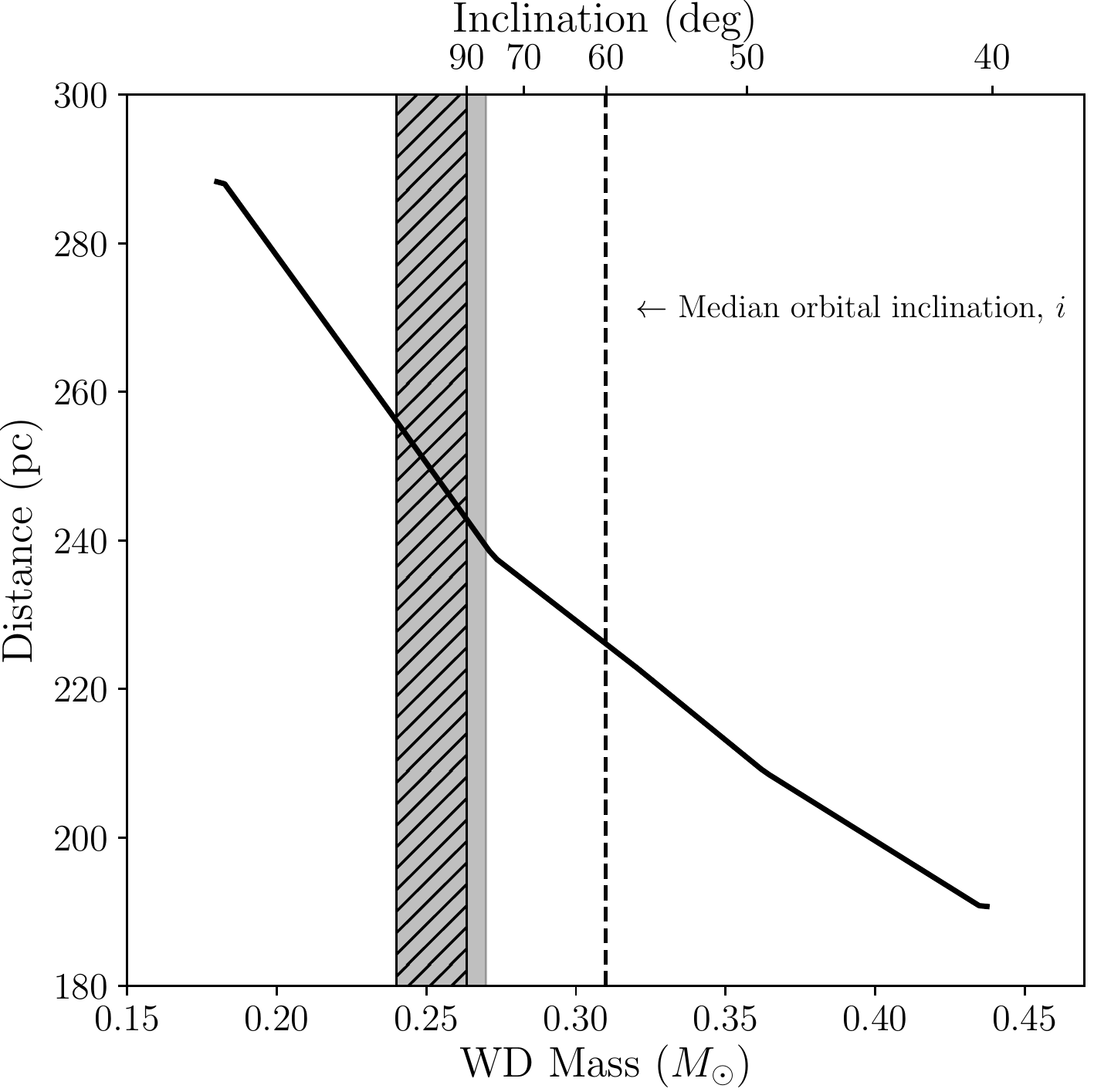}
    \caption{
      {\it Left:} Predicted distance as a function of intrinsic spin-down $\dot f$
      for \psr, based on Eqn.~\ref{eqn:varpi}.
      The red points show the 68\% confidence range for the
      distance posteriors. The grey band shows the inferred range 
      of distance for companion masses of $0.24-0.27$~\Msun, which 
      is the range inferred from Fig.~\ref{fig:PbMc}, based on the 
      models of low-mass white dwarfs from \cite{amc+13} computed for 
      $\Teff=3000\,$K and linearly interpolated for this mass range.
      We also show contours of the inferred X-ray and
      $\gamma$-ray efficiencies $\eta_{\rm X}$ and
      $\eta_\gamma$.  Values of $\eta_X$ between
     0.001\% and 0.01\% are consistent with the inferred
      distance range and are reasonable  given \citet{fhc+14},
      just as values of $\eta_\gamma$ between 1\% and 10\% are
      consistent with the data and are reasonable given
      \citet{gsl+16}.  Finally, the blue histogram shows the
      distribution of $\dot f$ for millisecond pulsars ($f>100\,$Hz)
      that are not in globular clusters based on \citet{mht+05}; they
      have been corrected for the Shklovskii effect as well as
      possible given the data in the catalog.  Note that the top $\dot
      P$ axis is only correct for a source with the spin period of
      \psr\ ($3.1\,$ms). {\it Right:} We show the same range of masses
      as indicated on the left-hand plot ($m_{\rm c}=0.24-0.27$~\Msun)
      and explicitly plot how distance scales with WD mass based on
      \cite{amc+13} models and our photometry results noted in \S\ref{sec:optical}.
      The top axis shows inclination angles corresponding to various WD
      masses and the hatched region shows WD masses excluded by $m_{\rm c, min}$;
      derived values assume $m_{\rm p}=1.35$~\Msun. 
    }
    \label{fig:distance_Fdot}
\end{figure*}

\section{Discussion} \label{sec:discuss}

With pulsar timing, we have measured \msp's parallax and find that $\dot{P}_{\rm meas}$ and $\mu_{\rm T}$ values place an upper limit on the pulsar's distance, which further constrains parallax and intrinsic spin-down. Combining these priors with another that accounts for the Lutz-Kelker bias, we find 95\% confidence intervals on parallax ($\varpi=3.7^{+1.6}_{-1.2}$~mas) and distance ($d=270^{+130}_{-80}$~pc) respectively. Furthermore, astrometric parameter measurements imply $\dot{P}_{\rm Shklov}=7(2)\times10^{-21}$, limiting intrinsic spin-down to $\dot{P}_{\rm int}\lesssim2.2\times10^{-21}$; only four other MSPs in the Galactic field (excluding those in globular clusters) have $\dot{P}_{\rm int}$ values this low \citep{mht+05}. For \msp this has interesting implications for other derived parameters such as characteristic age, $\tau>22$~Gyr. The fact that $\tau>\tau_{\rm Hubble}$ is not particularly concerning since it is well known that characteristic age derived in this fashion is a poor predictor of a recycled pulsar's true age \citep[e.g.][]{ctk+94,llf+95}. Using WD cooling models \citep{tbg+11,bwd+11}, we find more realistic cooling timescales, $5<\tau_{\rm cool}<9$~Gyr for assumed WD masses between $0.2-0.4$~\Msun. Since the WD is born as the recycling process concludes, $\tau_{\rm cool}$ is a better indicator of the system's true age. Assuming the true age of the pulsar is inside this range and magnetic dipole braking is entirely responsible for its spin-down (i.e. its braking index, $n=3$), \msp's post-recycling birth period was likely between $2.4-2.7$~ms, given a value of $\dot{P}_{\rm int}$ close to the limit shown in Table \ref{tab:1400par}. This result is insensitive to the choice of $n$; braking indices $1<n<3$ produce nearly identical ranges for birth period.

Since $\dot{P}_{\rm int}$ is proportional to the intrinsic spin-down luminosity ($\dot{E}_{\rm int}$), \msp's low $\dot{P}_{\rm int}$ value likely also affects its high-energy emission. Typically X-ray and $\gamma$-ray luminosities, $L_{\rm X}$ and $L_\gamma$, are expressed as a fraction of $\dot{E}_{\rm int}$ with corresponding efficiencies, $\eta_{\rm X}\equiv L_{\rm X}/\dot{E}$ and $\eta_\gamma\equiv L_\gamma/\dot{E}$; values for these efficiencies have been found in the ranges $0.001\%<\eta_{\rm X}<0.1\%$ \citep[see Figure 8 of][]{fhc+14} and $1\%<\eta_\gamma<100\%$ \citep{gsl+16}. Contours within these ranges are highlighted in Figure \ref{fig:distance_Fdot}. After correcting for the Shklovskii effect, \msp's spin-down luminosity is $\dot{E}_{\rm int}<3.0 \times 10^{33}$ erg s$^{-1}$ (see Table \ref{tab:1400par}).

Using a nominal distance of 270~pc and assuming a beaming factor $f_\Omega = 1$, the $\gamma$-ray luminosity $L_\gamma=4\pi f_\Omega d^2 F_\gamma=8.9\times 10^{31}$ erg s$^{-1}$ \citep[see Eq. 15 from][and description therein]{2PC}, where $F_\gamma$ is the measured $\gamma$-ray energy flux from Table \ref{tab:lat}. Based on the implied $\gamma$-ray efficiency of $\eta_\gamma\gtrsim3\%$ -- on the low-end of efficiencies found for MSPs in 2PC -- the pulsar produces plenty of energy to power the $\gamma$-ray source. We also note that $\dot{E}_{\rm int}/d^2 = 7.4 \times 10^{34}$ erg s$^{-1}$ kpc$^{-2}$, which is very high owing to the small distance. Over 75\% of radio MSPs with $\dot{E}_{\rm int}/d^2 > 1.5 \times 10^{34}$ erg s$^{-1}$ kpc$^{-2}$ have LAT-detected $\gamma$-ray pulsations \citep{gt+14}. Evidently, as observed from Earth, \msp\ is relatively inefficient at converting spin-down luminosity into $\gamma$-ray emission, and given the flux of the $\gamma$-ray emission, the modulation is more difficult to detect than for most other MSPs.

The X-ray luminosity of $1\times10^{29}$~ergs~s$^{-1}$ (0.3--10~keV; $d=270$~pc) makes \msp\ the least X-ray luminous rotation-powered MSP detected to date. For reference, it is more than an order of magnitude fainter than other nearby MSPs \--- PSRs J0437$-$4715, J2124$-$3358 \citep{zavlin+06}, and J0030+0451 \citep{bg+09} \--- all of which have luminosities of $10^{30}$~erg~s$^{-1}$ or higher. This striking difference can be attributed to \msp's much smaller spin-down luminosity ($\dot{E}$); the implied conversion efficiency from spin-down to X-ray luminosity for \msp\ is $\eta_{\rm X}>3.3\times10^{-5}$, consistent with $10^{-5}<\eta_{\rm X}<10^{-3}$ typically found for MSPs. On the other hand, if $\dot{E}_{\rm int}$ is close to the derived upper limit, the low X-ray luminosity might be an indication that the polar cap heating mechanism operates less efficiently in \msp for reasons that remain to be understood. 

\psr is in a nearly circular, 9.5~day orbit around its WD companion, which has a minimum mass of $m_{\rm c, min}=0.26$~\Msun (assuming $m_{\rm p}=1.35$~\Msun). Interestingly, this value is in remarkable agreement with the predicted $(P_{\rm b},m_{\rm WD})$-relationship (see Figure \ref{fig:PbMc}). The correlation between $P_{\rm b}$ and WD mass is an expected result of the relationship between the He-core mass and radius of a low-mass, red giant donor star, regardless of the mass present in its outer envelope \citep{savonije+87,ts+99}. Most WDs with measured masses follow this expected relationship (see Figure \ref{fig:PbMc}). The WD companion of PSR J1640+2224 is the most obvious exception, but was removed from Figure \ref{fig:PbMc} due to inconsistent conclusions about its mass based on pulsar timing and astrometric follow-up (S. Vigeland, private communication). Otherwise, only two $m_{\rm c,min}$ values are inconsistent with predicted curves.\footnote{These points correspond to PSRs J1125$-$6014 \citep{lfl+06} and J1748$-$2446W \citep{rhs+05}, but there is no mention in the literature of them being inconsistent with the expected $(P_{\rm b},m_{\rm WD})$-relationship.} \cite{imt+16} show that the $(P_{\rm b},m_{\rm WD})$-relationship has some width, depending on the metallicity of the progenitor of the WD companion. By allowing a range of $m_{\rm p}$, $i$ and WD progenitor metallicities for \msp's companion, we find a narrow range of $m_{\rm c}=0.24-0.27$~\Msun for $P_{\rm b}=9.5$~days (see Figure \ref{fig:PbMc}). The WD mass inferred from the $(P_{\rm b},m_{\rm WD})$-relationship is quite close to $m_{\rm c, min}$ (for $m_{\rm p}=1.35$~\Msun), suggesting the system is highly inclined.  However, there is considerable uncertainty in the $(P_{\rm b},m_{\rm WD})$-relation not only as a function of metallicity (as plotted) but due to the unknown history of the system, so it is also worth considering alternate constraints on the inclination.

\begin{figure*}[ht!]
\begin{center}
\includegraphics[width=0.9\textwidth]{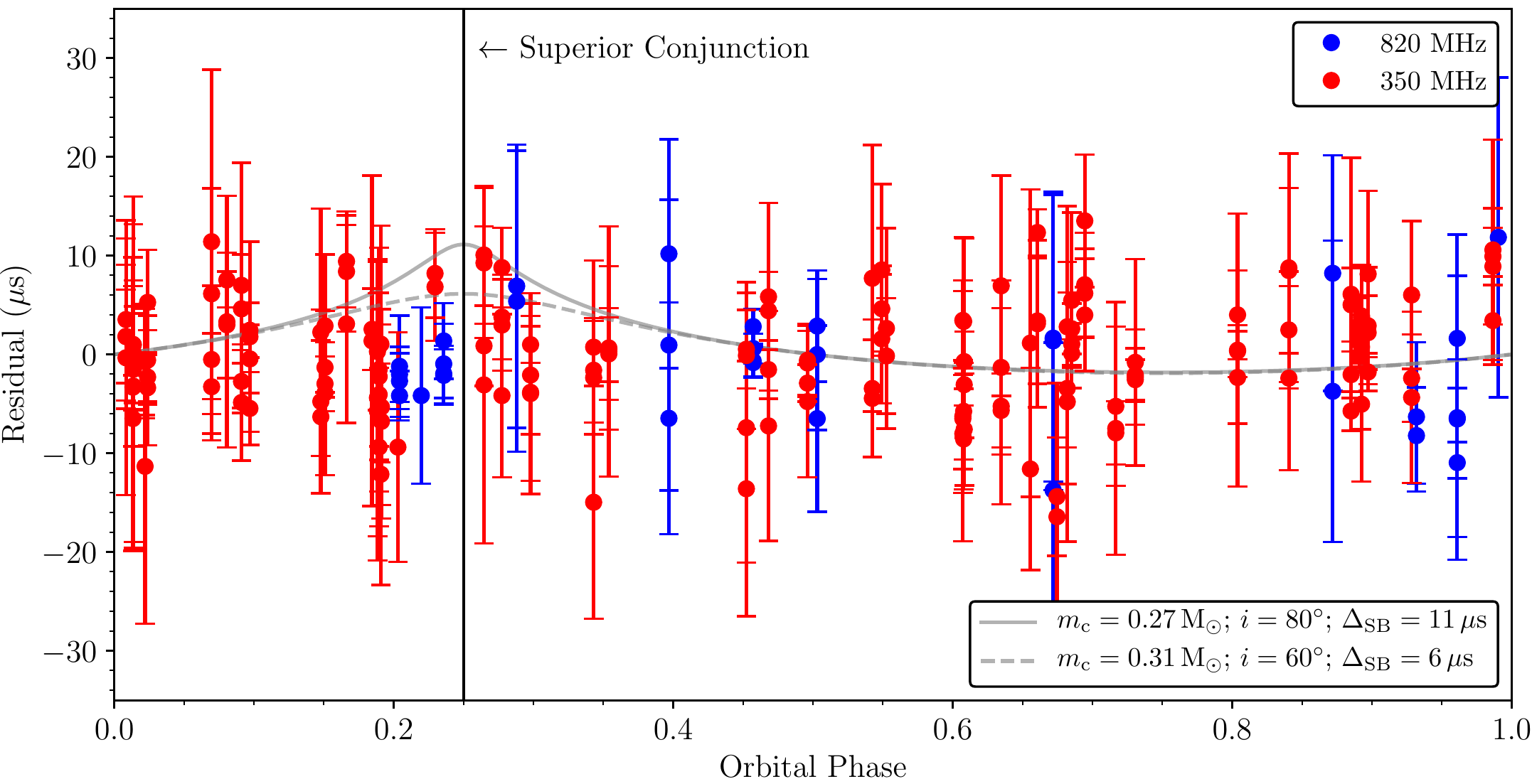}
\caption{Timing residuals in microseconds for \msp, plotted as a function of orbital phase. Observations at 350~MHz and 820~MHz are shown in red and blue, respectively. The solid/dashed gray lines show expected Shapiro delays, given assumed combinations of companion mass ($m_{\rm c}$) and orbital inclination angle ($i$). With these assumptions, the expected delays at superior conjunction ($\phi_{\rm orb}=0.25$) are 11~\us (solid) and 6~\us (dashed) respectively.}
\label{fig:orb_res}
\end{center}
\end{figure*}

For millisecond pulsars in highly-inclined orbits, a Shapiro delay signature is sometimes detectable in its timing residuals as a function of orbital phase.\footnote{We use {\it orbital phase} interchangeably with {\it eccentric anomaly}, since \msp's orbit is nearly circular.} The maximum delay occurs at superior conjunction (orbital phase, $\phi_{\rm orb}=0.25$), when the pulsar's signal must travel directly through its companion's gravitational well along our line of sight. If \msp were as highly inclined as discussed above, we would expect a Shapiro delay, $\Delta_{\rm SB}=11$~\us at superior conjunction (for $m_{\rm c}=0.27$~\Msun and $i=80^\circ$), which we do not see (Figure \ref{fig:orb_res}). However, going to the median expected inclination of $60\degr$ results in a qualitatively similar companion mass, $0.31$~\Msun, with a significant reduction in the Shapiro delay to 6~\us, which would not be detectable with the current data.  Note that a smaller pulsar mass could also reduce the significance of any Shapiro delay by moving to a lower implied inclination angle to match the $(P_{\rm b},m_{\rm WD})$-relation. We expect to be able to put better constraints on range and shape parameters after analyzing data from an upcoming, targeted Shapiro delay observing campaign.

We can further constrain the WD companion's mass using models \citep[e.g.][]{amc+13} that provide mass$-$radius relationships for low-mass WDs, photometry results from \S\ref{sec:optical} and the posterior PDF for distance (see Figure \ref{fig:distance}), derived from pulsar timing. Figure \ref{fig:distance_Fdot} (right panel) shows the conversion between WD mass and distance, which could further be expressed as a prior in $m_{\rm c}$-space; taking into account the low-significance parallax detection and additional priors mentioned in \S\ref{sec:dist_constraints}, a similar conversion effectively sets an upper limit on $m_{\rm c}\lesssim0.4$~\Msun.

Figure \ref{fig:distance_Fdot} (left panel) shows the remarkable agreement between the mass range predicted by the $(P_{\rm b},m_{\rm WD})$-relationship, our distance posterior (taking into account a significant $\dot{P}_{\rm Shklov}$), photometry results, and estimated X-ray and $\gamma$-ray efficiencies. The significant proper motion measured for \msp suggests that $\dot{P}_{\rm int}<2.2\times10^{-21}$, which is low, but still consistent with known values for other MSPs in the Galactic field whose intrinsic $\dot{P}$ values have been corrected for the Shklovskii effect. Figure  \ref{fig:distance_Fdot} (right panel) shows how the $m_{\rm c}=0.24-0.27$~\Msun range is mostly excluded, simply based on $m_{\rm c, min}$ (assuming $m_{\rm p}=1.35$~\Msun), derived from timing results. The lack of detectable Shapiro delay implies a slightly higher companion mass and lower inclination angle. Despite slight inconsistency with the mass range implied by the $(P_{\rm b},m_{\rm WD})$-relationship, our data suggest \msp's companion mass is likely $\sim0.30$~\Msun and the system is $\approx230$~pc away with an orbital inclination angle, $i\gtrsim60^{\circ}$.

\section{Conclusions} \label{sec:conclude}
In this paper, we described follow-up timing efforts on \psr since its discovery by high school students involved in the Pulsar Search Collaboratory was first reported in \cite{rsm+13}. Our updated solution includes TOAs spanning five years from timing observations conducted with the GBT at $\sim$monthly cadence. With the latest timing solution, we measure \msp's position to milli-arcsecond precision, its spin-down, proper motion, a monotonic slope in DM over time, and a weak parallax signature. Because of the pulsar's significant total proper motion, a kinematic (Shklovskii) component accounts for a significant fraction of $\dot{P}_{\rm meas}$ and we can only place an upper limit on the intrinsic spin-down, $\dot{P}_{\rm int}<2.2\times10^{-21}$~s/s. 

The Shklovskii effect provides an additional prior for the system's parallax and in turn, better constraints on distance, $d=270^{+130}_{-80}$~pc. This range agrees nicely with distances estimated using electron density models \citep[$270-500$~pc; ][]{tc+93, cl+02, ymw+17}.

Using the Goodman Spectrograph on the 4.1-m SOAR Telescope and later, the LRIS on the 10-m Keck I Telescope for deeper imaging, we detected \msp's WD companion for the first time. Photometry suggests the companion is a cool, DA-type WD (Hydrogen atmosphere) with $T_{\rm eff}=3000\pm100$~K and $R/R_\odot=(2.19\pm0.03)\times10^{-2}\,(d/270~{\rm pc})$. Combined with WD cooling models, the effective temperature measurement suggests that the system's age is in the range $5-9$~Gyr, which is consistent with the relatively low upper limit we place on $\dot{P}_{\rm int}$ after correcting for the Shklovskii effect and the corresponding characteristic age. Using WD mass-radius models from \cite{amc+13} and photometric $R/d$, we find implied mass and distance ranges completely consistent with $m_{\rm c,min}=0.26$~\Msun and $d=270^{+130}_{-80}$~pc measurements.

Finally, with high-energy detections of \msp with \xmm and \fermi, we measured X-ray and $\gamma$-ray luminosities, $L_{\rm X}=1\times10^{29}$~ergs~s$^{-1}$ and $L_\gamma=8.9\times10^{31}$~ergs~s$^{-1}$, respectively. Given the upper limit on $\dot{P}_{\rm int}$ (and therefore $\dot{E}_{\rm int}$), we find efficiencies $\eta_{\rm X}>3.3\times10^{-5}$ and $\eta_\gamma\gtrsim0.03$, consistent with expected ranges for respective wavelength regimes. Although measured high-energy luminosities depend on the assumed nominal distance ($d=270$~pc), corresponding efficiencies provide additional consistency checks on $\dot{P}_{\rm int}$, distance and photometry constraints determined with various methods.

This information presents a consistent picture; combined, it suggests \psr has an intrinsic spin-down $\dot{P}_{\rm int}\approx2\times10^{-21}$~s/s, a distance $d\approx230$~pc, WD companion mass $m_{\rm c}\sim0.30$~\Msun, and orbital inclination $i\gtrsim60^{\circ}$. These conclusions are slightly inconsistent with WD evolution models \citep[e.g.][]{imt+16} and depend on an assumed pulsar mass ($m_{\rm p}=1.35$~\Msun), but our results are relatively insensitive to $m_{\rm p}$. Even for low orbital inclination angles ($i\sim60^{\circ}$), we expect a Shapiro delay signature to be detectable ($\Delta_{\rm SB}=6$~\us) with data from an upcoming, targeted observing campaign, which will provide further clarity on results presented here.

\section*{Acknowledgments}
The Green Bank Observatory is a facility of the National Science Foundation operated under cooperative agreement by Associated Universities, Inc.

JKS, DLK, MAM, DRL, PSR, RL, PG, and KS are supported by the NANOGrav NSF Physics Frontiers Center award number 1430284. Portions of this research performed at the Naval Research Laboratory are supported by NASA. RH, AV, PC, and BB would like to thank the High Point University (HPU) Student Government Association for providing travel funds to Cerro Tololo/Pachon; the HPU Summer Undergraduate Research Program in the Sciences for providing summer research support; and President Qubein, Provost Carroll, and Dean Stoneking for their generous support of the sciences at HPU. AGI acknowledges support from the NASA Astrophysics Theory Program through NASA grant NNX13AH43G.

We thank Matthew Kerr for providing the 2PC H-test values.
The \fermi\ LAT Collaboration acknowledges generous ongoing support
from a number of agencies and institutes that have supported both the
development and the operation of the LAT as well as scientific data
analysis.  These include the National Aeronautics and Space
Administration and the Department of Energy in the United States, the
Commissariat \`a l'Energie Atomique and the Centre National de la
Recherche Scientifique / Institut National de Physique Nucl\'eaire et
de Physique des Particules in France, the Agenzia Spaziale Italiana
and the Istituto Nazionale di Fisica Nucleare in Italy, the Ministry
of Education, Culture, Sports, Science and Technology (MEXT), High
Energy Accelerator Research Organization (KEK) and Japan Aerospace
Exploration Agency (JAXA) in Japan, and the K.~A.~Wallenberg
Foundation, the Swedish Research Council and the Swedish National
Space Board in Sweden. Additional support for science analysis during the operations phase is
gratefully acknowledged from the Istituto Nazionale di Astrofisica in
Italy and the Centre National d'\'Etudes Spatiales in France.

A portion of the results presented was based on observations obtained with \textit{XMM-Newton}, an ESA science mission with instruments and contributions directly funded by ESA member states and NASA.

This paper is based (in part) on data obtained with the International LOFAR Telescope (ILT). LOFAR \citep{vwg+13} is the Low Frequency Array designed and constructed by ASTRON. It has facilities in several countries, that are owned by various parties (each with their own funding sources), and that are collectively operated by the ILT foundation under a joint scientific policy.

{\it Facilities:} GBT (GUPPI), \fermi LAT, \xmm (pn, MOS1/2), Keck I: 10-m (LRIS), SOAR: 4.1-m (Goodman Spectrograph), LOFAR, LWA

{\it Software:} libstempo, astropy \citep{astropy}, scipy \citep{scipy}, {\sc Tempo}, {\sc Tempo2} \citep{tempo2}, {\sc PSRCHIVE} \citep{psrchive}, IRAF/DAOPHOT, {\sc LPipe}, {\tt SExtractor} \citep{ba+96}, {\it Fermi} Science Tools


\end{document}